\documentclass[preprint2]{aastex}

\shorttitle{NGC~602 SFH}
\shortauthors{Cignoni et al.}

\begin{document}


\title{Star formation history in the SMC: the case of NGC~602\altaffilmark{1}}

\author{M. Cignoni\altaffilmark{2,3}, E. Sabbi\altaffilmark{4,5}, A. 
Nota\altaffilmark{4,5}, M. Tosi\altaffilmark{3}, 
S. Degl'Innocenti\altaffilmark{6,7}, P.G. Prada Moroni\altaffilmark{6,7}, 
L. Angeretti\altaffilmark{3}, Lynn Redding Carlson\altaffilmark{8}, J. 
Gallagher\altaffilmark{9}, M. Meixner\altaffilmark{4}, 
M. Sirianni\altaffilmark{4,5}, L.J. Smith\altaffilmark{4,5,10}} 

\altaffiltext{1}{ Based on observations with the NASA/ESA Hubble Space
Telescope, obtained at the Space Telescope Science Institute, which is
operated by the Association of Universities for Research in Astronomy (AURA),
Inc., under NASA contract NAS5-26555. These observations are associated with program \#10248.}

\altaffiltext{2}{Dipartimento di Astronomia, Universit\`a degli Studi di
Bologna, via Ranzani 1, I-40127 Bologna, Italy}

\altaffiltext{3}{Istituto Nazionale di Astrofisica, Osservatorio Astronomico
     di Bologna, Via Ranzani 1, I-40127 Bologna, Italy}

\altaffiltext{4}{Space Telescope Science Institute, 3700 San Martin Drive,
Baltimore, USA}

\altaffiltext{5}{European Space Agency, Research and Scientific Support
Department, Baltimore, USA} 
\altaffiltext{6}{Dipartimento di Fisica ``Enrico Fermi'', Universit\`a di Pisa, largo Pontecorvo 3, Pisa I-56127, Italy} 
\altaffiltext{7}{INFN - Sezione di Pisa, largo Pontecorvo 3, Pisa I-56127,
   Italy}
\altaffiltext{8}{Department of Physics and Astronomy, Johns Hopkins University, Baltimore, Usa}
\altaffiltext{9}{University of Wisconsin, Madison, WI, USA}
\altaffiltext{10}{Department of Physics and Astronomy, University College
London, Gower Street, London, UK}

\begin{abstract}

Deep HST/ACS photometry of the young cluster NGC 602, located in the remote low density "wing" of the Small Magellanic Cloud, reveals numerous pre-main sequence stars as well as young stars on the main sequence. The resolved stellar content thus provides a basis for studying the star formation history into recent times and constraining several stellar population properties, such as the present day mass function, the initial mass function and the binary fraction. To better characterize the pre-main sequence population, we present a new set of model stellar evolutionary tracks for this evolutionary phase with metallicity appropriate for the Small Magellanic Cloud (Z = 0.004). We use a stellar population synthesis code, which takes into account a full range of stellar evolution phases to derive our best estimate for the star formation history in the region by comparing observed and synthetic color-magnitude diagrams. The derived present day mass function for NGC~602 is consistent with that resulting from the synthetic diagrams. The star formation rate in the region has increased with time on a scale of tens of Myr, reaching $0.3-0.7 \times 10^{-3} M_\odot yr^{-1}$ in the last 2.5 Myr, comparable to what is found in Galactic OB associations. Star formation is most complete in the main cluster but continues at moderate levels in the gas-rich periphery of the nebula.
\end{abstract}


\section{Introduction} 

The Small Magellanic Cloud (SMC), the closest dwarf irregular galaxy, is the most appropriate target for detailed studies of resolved stellar populations
in this extremely common class of objects. Its present-day metallicity
($Z=0.004$) makes the SMC the best local analog to the vast majority
of late-type dwarfs. Deep optical photometry was performed on images
from the Hubble Space Telescope (HST) Advanced Camera for Surveys
(ACS) at different locations in the SMC as part of a program to study
the global Star Formation History (SFH) of this galaxy and to
understand how star formation is affected by local conditions such as
metallicity and dust content. In the present paper, we derive the SFH
for NGC~602, a very young cluster located in the ``wing'' of SMC. The
NGC~602 region offers a valuable laboratory to study infant stars (1-3
Myr) as well as billion year old objects. In particular, \citet{carlson07} identified a rich population of pre-main sequence (PMS) stars and candidate Young Stellar Objects (YSOs) using the Color-Magnitude Diagrams (CMDs) obtained from the HST optical data and Spitzer Space Telescope IR data. There is no question that the well defined stellar sequence, redder than the canonical main sequence is due to PMS stars.  Conventional alternative explanations (reddening, binaries, photometric errors) have been taken into account but discarded because they fail to reproduce the CMD feature.

The discovery of low mass PMS stars in nearby
galaxies \citep[see also][]{roma,nota06,gouliermis07} is important for
at least two reasons. First, it is an excellent benchmark for stellar
evolution calculations, since it provides the opportunity to study
rather large samples of PMS stars characterized by similar distance
and chemical composition.  Second, the detection of PMS stars with
precise photometry gives an additional channel, with respect to the
main sequence, to constrain the recent SFH: the short time spent in
PMS by a $1\,M_{\odot}$ star ($\sim 50$ Myr) roughly corresponds to
the time spent by an $8\,M_{\odot}$ star on the main sequence, with
the obvious advantage that stars of $1\,M_{\odot}$ are much more
frequent than $8\,M_{\odot}$ stars.

The goal of this paper is to recover the SFH of NGC~602. For this task, we
merged a well tested synthetic population code \citep{tosi} with an effective
maximum likelihood technique \citep[see e.g.][]{cigno06}, to build accurate
theoretical CMDs to be compared with the observations.  The synthetic CMDs are
generated by Monte Carlo sampling of canonical ingredients such as a star
formation law, initial mass function (IMF), age-metallicity relation (AMR) and
stellar evolution tracks.  In order to properly account for the PMS phase, new
PMS tracks at $Z=0.004$ have been calculated.

This paper is organized as follows. In section 2 the data are briefly
described.  In Section 3 we present the new set of PMS models. In Section 4 we
disentangle the observed CMD in terms of age, metallicity, distance and
reddening. In addition, the present day mass function (PDMF) is derived and
some hypotheses on the interplay of the star formation rate (SFR) with the environment are done. Section 5 is
devoted to the method and the stellar ingredients we adopt for reconstructing
the SFH.  Finally, in section 6 we present our main quantitative results on
the SFH.

\section{Observations}
For this project, we utilized observations of NGC~602 (R.A.$=01^h 29^m 31^s$,
Dec$=-73\degr 33\arcmin 15\arcsec$, J2000) taken with the Wide Field Channel
(WFC) of the HST/ACS as part of proposal 10248 (P.I. A. Nota).  Five long
exposures were taken through filters F555W ($\sim$V) and F814W ($\sim$I). A
dither pattern was especially designed to allow for hot pixel removal, and to
fill the gap between the two halves of the $4096\times 4096$ pixel detector.
This dithering technique also improved both point spread function (PSF)
sampling and photometric accuracy by averaging flat-field errors and smoothing
over the spatial variations in detector response. All the images were taken
with a gain of 2 e$^-$ ADU$^{-1}$. The entire data set was processed with the
CALACS calibration pipeline, and for each filter the long exposures were
co-added using the multi-drizzle package. The final exposure times of the deep
combined F555W and F814W are $2150\, {\rm s}$ and $2265\, {\rm s}$
respectively. The image covers a region of $200\arcsec\times 200\arcsec$
corresponding to a linear size of $58\times 58\, {\rm pc^2}$ at the SMC distance
\citep[see Fig. 1 in][]{carlson07}.

Additional short B, V and I ground-based observations were taken with the
camera ANDICAM on the 1.3 m Small and Moderate Aperture Research Telescope
System (SMARTS), to recover the photometric information of the brightest stars,
that are heavily saturated in the ACS data. Photometric observations included
three 60 second B-band exposures, three 45 second V-exposures, and three 30
second I-exposures.

\begin{figure}
\plotone{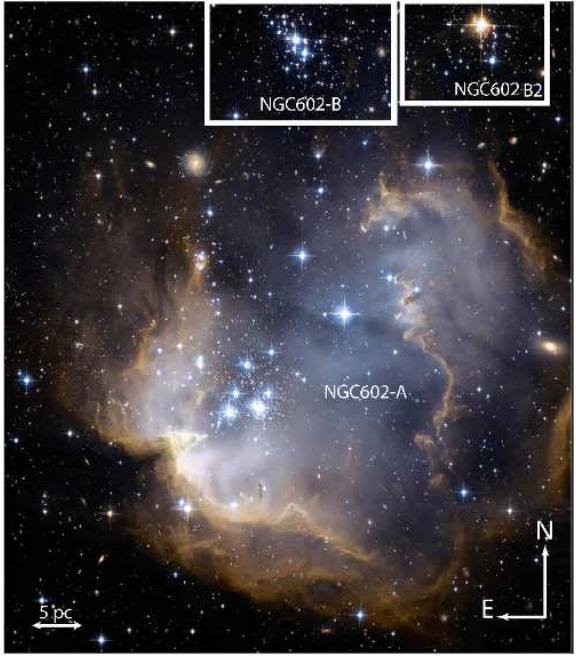}
\caption{\label{xy} ACS image for the observed region around NGC~602. The sub-clusters
NGC~602-B and NGC~602-B2 are also indicated.}
\end{figure}

Figure \ref{xy} shows the ACS image of the region. In addition to the central cluster, two other sub-clusters
are identified in the region, clearly separated from the central cluster.
Following the nomenclature in \citet{west}, we call the central cluster
NGC~602-A and the top-central sub-cluster NGC~602-B. The third sub-cluster is
thereinafter called NGC~602-B2 (corresponding to ``cluster A'' in
\citealt{schma}).

\begin{figure}
\plotone{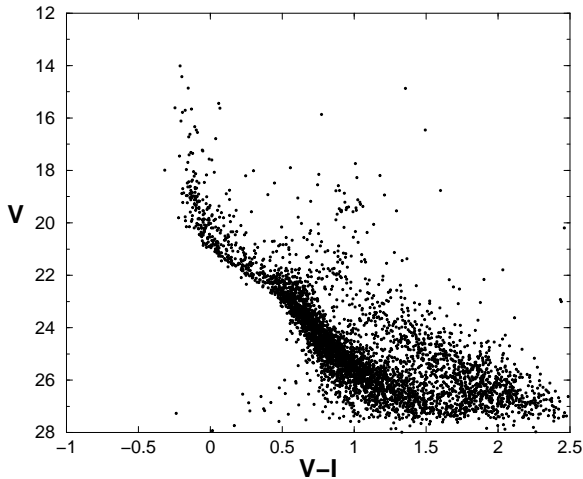}
\caption{\label{fig0} 
Optical Color Magnitude Diagram for the entire NGC~602  region.}
\end{figure}

As part of the same program we observed NGC~602 core with the Near Infrared Camera and Multi Object Spectrometer (NICMOS). We used ACS in parallel to observe with the WFC a SMC reference field ($\alpha=01^h29^m19^s; \delta=-73\degr 42\arcmin 30.8\arcsec$, J$_{2000}$). The reference field is located $\sim 9\arcmin 46\arcsec$ ($\sim 172$ pc) south from NGC~602. The data were analyzed following the same procedure described above, and used to subtract the contamination from the SMC field in the study of the present day mass function (PDMF, \S~\ref{MF}). 
 
\subsection{Photometric Reduction}

The photometric reduction has been performed with the DAOPHOT package within
the IRAF\footnote{IRAF is distributed by the National Optical Observatories,
which are operated by AURA Inc., under cooperative agreement with the National
Science Foundation.} environment.

We have followed the same method used for NGC~330 \citep{sirianni02} and
NGC~346 \citep{sabbi07}. To derive accurate photometry of all stars in the
field, we adopted the PSF fitting and aperture photometry routines provided
within DAOPHOT.

The stars in each image were independently detected using DAOFIND. We set the
detection threshold to $4 \sigma$ above the background level in the F555W
image and $3.5 \sigma$ in the F814W. We measured the stellar flux by aperture
photometry with PHOT, and then refined the photometric measurements of the
individual sources with PSF-fitting photometry. For both the F555W and F814W
images we computed spatially variable PSFs by selecting more than 170 isolated
and moderately bright stars, uniformly distributed over the images, in order
to take into account the variations in shape and in core width of the PSF
across the images \citep{krist03,sirianni05}. Stars with $m_{F555W} <18$ and
$m_{F814W} <17.5$ were discarded from the long exposure photometric catalog
because they were saturated.

Many background galaxies are visible in the HST/ACS images \citep{nigra}.  For a safe interpretation of the characteristics of the observed stellar populations, we needed to distinguish true single stars from extended, blended and/or residual spurious objects such as background galaxies. Thus, we applied to our catalogs selection criteria based on the shape of the objects and on the quality of photometry. We considered the DAOPHOT parameters $\chi^2$, sharpness and
photometric error $\sigma_{\rm DAO}$. $\chi^2$ gives the ratio of the observed
pixel-to-pixel scatter in the fit residuals to the expected scatter calculated
from a predicted model based on the measured detector features. The sharpness
parameter sets the intrinsic angular size of the objects.  We selected only
the objects having $-0.5<sharpness_{m_{F555W}}<0.5$ and
$-0.3<sharpness_{m_{F814W}}<0.3$. These values turned out to be the best ones
to allow us to reject background galaxies and fuzzy brightness variations in
diffuse emission regions. Our HST catalog includes 4496 sources.

We used ground-based short V and I exposures to recover the photometric
information of the saturated stars. We used the IRAF packages XREGISTER,
IMSHIFT, and IMCOMBINE to properly align and combine the images in each
band. We used DAOPHOT's DAOFIND, with the detection threshold set at $4
\sigma$ from the background, to construct a photometry list. Stellar fluxes
were measured by aperture photometry with an aperture size of 4 times the PSF
FWHM, and then we refined the photometry by PSF-fitting. We used Landolt stars
to calibrate the ground-based photometry into the Johnson-Cousins photometric
system, and then we followed the prescription by \cite{sirianni05} to convert
the photometric catalog into the Vegamag photometric system. The entire
optical catalog includes 4537 stars calibrated in the HST Vegamag system. The resulting CMD is presented in Figure~\ref{fig0}.

\subsection{The Photometric Completeness}

Reliable quantitative derivations of the evolutionary properties of the observed region require robust estimates of the photometric errors and incompleteness of our images. Spreads in the CMD sequences due to photometric errors could in fact mimic age, distance or reddening effects, and incompleteness competes with the SFH and the IMF in defining the observed luminosity functions. 

In order to test the level of completeness of our photometric data and to have
a more reliable estimate of the photometric errors and blending effects on our
data, we run a series of extensive artificial star experiments, following the
same procedure described in \cite{sabbi07b}. The experiments consist of adding
``artificial'' stars, obtained from the scaled PSF used in the photometric
analysis of the frames, onto the actual image. Then the artificial stars are
retrieved using exactly the same procedure adopted for the data reduction. The
completeness is defined as the ratio between the stars retrieved and the stars
added. In total, more than 1,000,000 artificial stars were simulated in the
F555W and F814W deep exposures. The magnitudes of the artificial stars were
chosen to mimic the observed distribution of stars in the CMD. Regions with no
observed stars were assigned a minimum number of artificial stars to fully
cover the CMD.  We placed the artificial stars one at a time on a regular grid
such that the spacing between them was 5.3 PSF-fitting radii (80 pixel), to
ensure that the artificial stars experiments did not change the crowding
conditions of the data. To fully sample the completeness level of the data,
the absolute position of the grid with respect to the frame was randomly
changed at each run.

\begin{figure}
\plotone{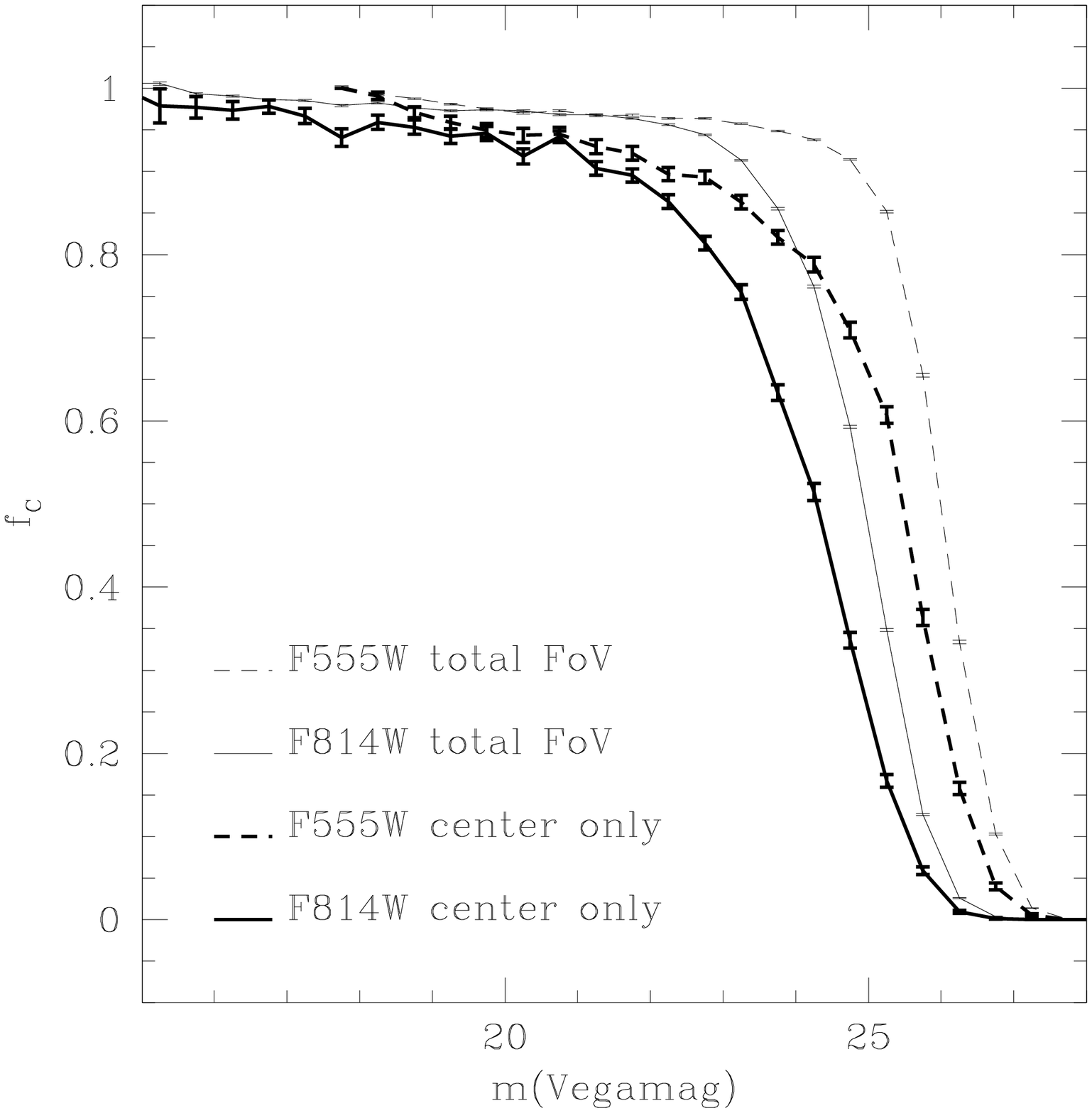}
\caption{Completeness curves as derived from the artificial star experiment for
the F555W ({\it dashed line}) and F814W ({\it solid line}) photometry as a
function of magnitude. Completeness curves have been derived both for the entire field of view (narrow lines) and for the central region only (thick lines).}
\label{f:compl} 
\end{figure}

The frame to which the artificial stars were added was reduced exactly as the
original frame. We determined the level of photometric completeness by
comparing the list of artificial stars ``added'' with the list of the stars
``recovered''.  In our experiments each star found contemporarily in both
frames with an input-output magnitude difference $\Delta m\le 0.75$ and
satisfying the selection criteria listed in \S 2.1 is considered a ``recovered''
star.  The completeness factor as a function of magnitude for both filters
F555W and F814W is presented in Figure~\ref{f:compl}. This completeness factor and the photometric errors derived here have been adopted to create the synthetic CMDs described in Sect.5. 

\section{Stellar tracks}
The synthetic CMDs used to reconstruct the SFH of NGC~602 are calculated
adopting suitable theoretical stellar tracks, from the PMS birth-line to the
AGB phase. For all the stellar phases except the PMS, we use the Padua
evolutionary tracks \citep{fagotto94}. The PMS tracks are new and specifically
computed with an updated version of the FRANEC evolutionary code \citep[see
e.g.][]{chieffi,deg07}. For these models the OPAL 2006 equation of state
(EOS)\footnote{http://www-phys.llnl.gov/Research/OPAL/EOS\_2005/} has been
adopted \citep[see also][]{ig96}, together with radiative opacity tables by
the Livermore
group\footnote{http://www-phys.llnl.gov/Research/OPAL/EOS\_2005/} \citep[see
also][]{ig96} for temperatures higher than 12000$^{o}$K; in this way the EOS
and opacity calculations are fully consistent. The conductive opacities are
from \cite{shtern} \cite[see also][]{pot} while the atmospheric opacities are
from \cite{ferg2005}. All the opacity tables have been calculated adopting the
\cite{aspl} solar mixture. Nuclear reaction rates are from the NACRE
compilation \citep{angu}.

As is well known, the treatment of the early PMS phases is very problematic
and the definition of the birth-line (that is the phase in which the
proto-star reaches thermal and hydrostatic equilibrium and can be thus
represented in a CMD) requires the adoption of hydrodynamical codes \citep[see
e.g.][for an extensive analysis]{wuch,stah}.  In addition, the results are
influenced by several physical parameters still not well known \citep[the
accretion rate, the deuterium mass fraction in the interstellar medium, and
the stellar opacity, see e.g.][]{berna} and thus the analysis can be only
semi-quantitative. As a general procedure in stellar evolutionary codes, our
calculations artificially start from a total convective model at the
hydrostatic and thermal equilibrium. However this should not constitute a
problem: numerical simulations have already shown that the stellar
characteristics after the birth-line are not significantly influenced by the
previous evolutionary history. The birth-line is supposed to be reached very
early (our synthetic CMDs are built with a birth-line at 0.5 Myr, which is
fairly compatible with the data CMD) and thus, at least for masses that are
not too high ($<5-6\,M_{\odot}$), the greater part of the PMS evolution can be
represented in the CMD.

\begin{figure}
\plotone{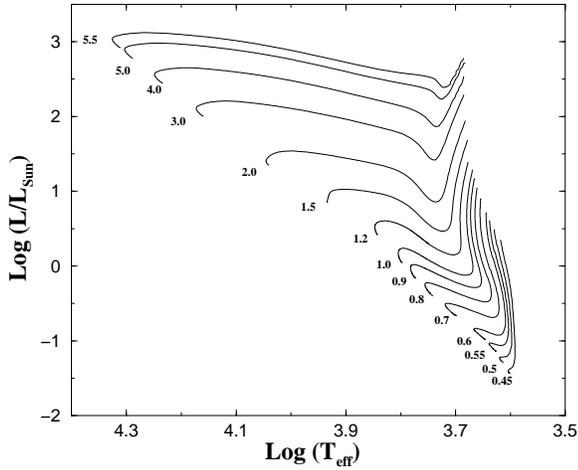}
\caption{PMS evolutionary tracks for $Z=0.004$, where the  characteristic mass
for each track is given, in solar masses.}
\label{figLT} 
\end{figure}

Models have been calculated with Z=0.004 Y=0.24 and the mixing-length
parameter $\alpha=$1.9 for masses in the range $0.45-5.5\,M_{\odot}$ (all PMS
models are presented in Table \ref{tab1}). Fig.~\ref{figLT}
shows our computed PMS tracks in the theoretical Luminosity-Temperature
plane. 

\begin{figure}
\plotone{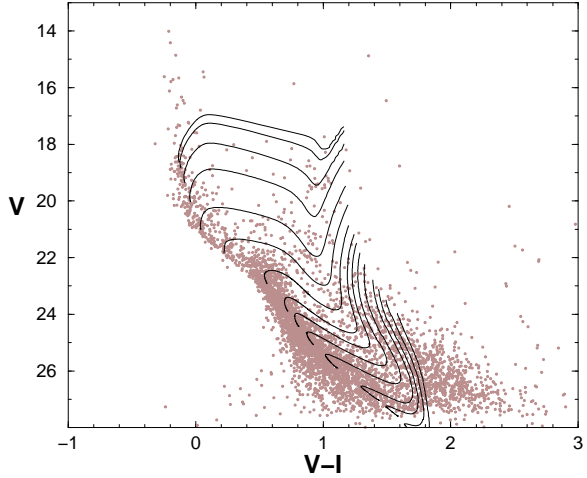}
\caption{FRANEC PMS stellar tracks ($Z=0.004$) overlaid on the observed CMD.}
\label{fig01} 
\end{figure}

To convert all stellar tracks (Padua and Pisa) from the
Luminosity-Temperature plane to the CMD we used the transformations by
\cite{origlia00} for the HST Vegamag photometric system. The transformed PMS
tracks are shown in Fig. \ref{fig01}.

\section{General considerations from the CMD analysis}

Although the most striking feature is the presence of a wealth of PMS
stars, the CMD also shows a clear signature of an old population. In
particular, we see a main sequence (MS) turn-off (TO) near
$M_{F555W}\sim 22$, a sub-giant branch (SGB) also at $M_{F555W}\sim
22$, and a red clump (RC) around $M_{F555W}\sim 19.5$. Excluding a large contamination by PMS stars more massive than $1.5\, M_\odot$, this is likely die to the SMC field population, still present also in these low-density ``wing'' areas. A contamination of the SGB by high mass
PMS stars cannot be ruled out.  From the evolutionary point of
view, the presence of PMS stars indicates a SF activity in the last
few million years, while the width of the upper main sequence suggests
an overall activity in the last 100 Myr. The RC represents a
population older than 500 Myr burning helium in the stellar core and
the SGB at that luminosity level indicates that stars of several Gyr
are present in our sample - likely the SMC field population.

\begin{figure}
\plotone{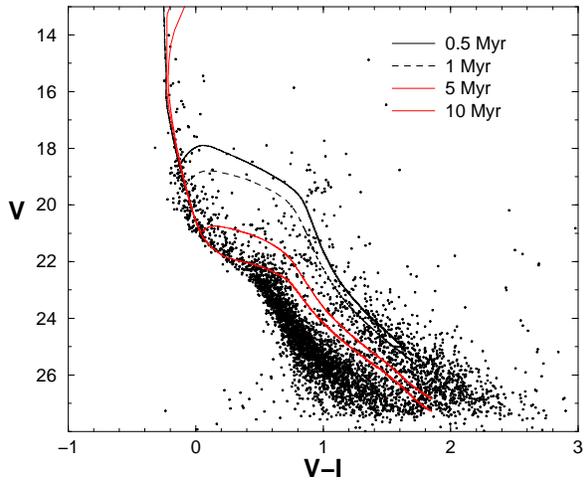}
\caption{0.5, 1.0, 5.0 and 10.0 Myr isochrones for metallicity $Z=0.004$
interpolated from the Padua tracks and FRANEC tracks for PMS overlaid on the
observed CMD. Distance modulus and reddening are assumed equal to
$(m-M)_0=18.9$ and $E(B-V)\sim 0.08$, respectively.}
\label{fig02} 
\end{figure}

Fig.~\ref{fig02} shows isochrone fitting of the young stellar
population using Padua stellar models for Z=0.004, corrected for
distance ${(m-M)_0}=18.9$ and reddening $E(B-V)\sim 0.08$.  In the
magnitude range $V=15-19$ a stellar excess is observed on the red side
of the isochrones. We exclude the possibility that this is a pure age
effect. In fact, although aging the youngest stars by some Myr would
move the bright end of the main sequence to the red, it would have
also the collateral effect of producing an unobserved blue-shift of
the PMS population. We consider it more likely that our sample be
affected by significant differential reddening, with some of the
brighter stars more reddened than the faint ones. If these stars are
actually the massive counterpart of the observed faint PMS, these
stars can be still surrounded by relics of their birthing cocoon
material and suffer an additional amount of absorption. Another
possibility is that some of the most massive stars are rotating. In
this case, these stars would be shifted to lower effective
temperatures as a consequence of the lower internal pressure with
respect to non-rotating ones (see e.g. \citealt{mey}).

Besides the upper main sequence, the PMS population gives constraints on our
age estimates. Fig. \ref{fig01} shows the observed CMD with the PMS tracks
(masses from 0.45 to 5.5 $M_\odot$) for metallicity $Z=0.004$ over-plotted. It
is clear that the theoretical tracks comfortably encompass the red sequence,
confirming the pre-main sequence nature of these sources. Even more
interesting, the bulk of these stars seem to lie in the Hayashi region. Figure
\ref{fig02} shows the corresponding PMS isochrones for 0.5 Myr, 5 Myr and 10
Myr, and suggests that most of the PMS stars are younger than 5 Myr.  The same
Figure underlines the effectiveness of the PMS stars as astrophysical clocks:
the upper main sequence alone is degenerate with respect to progressive epochs
of recent star formation, and it is impossible to resolve isochrones of
different ages. In contrast, the same epochs of star formation are clearly
separated in the PMS region that, therefore, offers a more accurate snapshot
of the star formation activity in the last 50 Myr.

A further inspection of the PMS region reveals that, at least for $V>23$, many
objects are up to 1 magnitude redder than our PMS tracks. The number of these
red stars is too large to ascribe them to photometric errors. More likely some
very young objects continue to have significant amounts of dust, thus
suffering additional self-absorption (see also Section 6.1.4 for a discussion).

\begin{figure}
\plotone{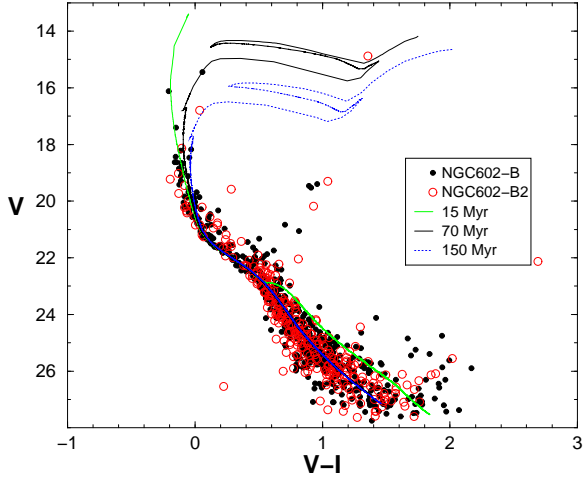}
\caption{Isochrones overlaid on the CMD for the regions of  NGC~602-B and
NGC~602-B2.}
\label{2clusters} 
\end{figure}

One additional question is whether the sub-clusters NGC~602-B and
NGC~602-B2 have a different history to the central active region
(NGC~602-A). Figure \ref{2clusters} shows the CMD for the stars found
in the boxes\footnote{Although the shape of these selection areas
may lead to overestimate the field stars, the aim here is only to
investigate the stars around the two overdensities (whose boundaries
may be quite irregular).} of Figure \ref{xy}, chosen to represent
NGC~602-B and NGC~602-B2. Both clusters are reproduced using
isochrones with ages between 15 and 150 Myr, slightly older than
NGC~602 ($\sim$ 10 Myr). It must be emphasized however that this
difference in age is due to only one and two stars evolved off the MS
in NGC~602-B and NGC~602-B2 respectively. Due to the low number of
stars the sub-clusters NGC~602-B and NGC~602-B2 are indistinguishable
from NGC~602-A. In other words, the same star formation process that
took place on various scales in the whole region spawned also these
two small associations.

Looking for ancient populations, Figure \ref{2clusters} shows also that a
well-defined SGB along with a RC is a common feature of the CMD for NGC~602-A,
B and B2. This finding is a robust indicator of an old population, being the
SGB and the RC phases populated by low mass stars. Moreover, the compact
morphology of the RC suggests that the metallicity does not vary in the sample
in a relevant way.

\begin{figure}
\plotone{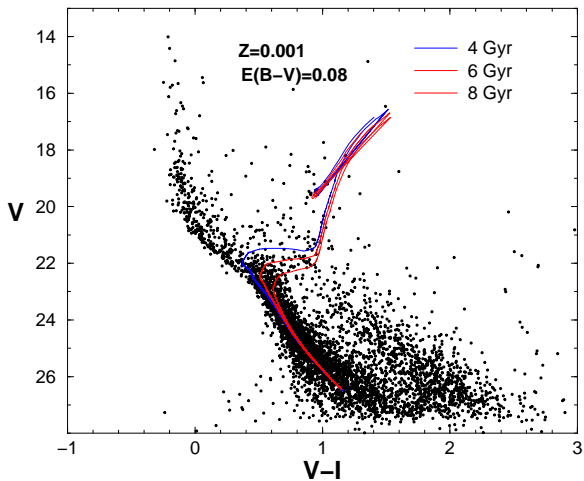}
\caption{4.0, 6.0 and 8.0 Gyr Padua isochrones for $Z=0.001$
(\citealt[see][]{ange07}) at the distance modulus $(m-M)_0=18.9$ and reddening
$E(B-V)=0.08$ superimposed on the observed CMD.}
\label{fig03} 
\end{figure}

By means of isochrone fitting, we find that luminosities and colors of RC and
SGB can be simultaneously reproduced by assuming a metallicity of $Z=0.001$, a
distance modulus ${(m-M)_0}=18.9$, reddening $E(B-V)\sim 0.08$, and an age
between 4 and 8 Gyr (see Fig.~\ref{fig03}).

\subsection{The Present  Day  Mass Function}
\label{MF}

From an observational point of view, the mass function (MF) is the fractional
number of stars per mass interval per unit area. Here we adopt the
parameterization by \cite{scalo86}, in which the mass function is
characterized by a slope $\Gamma=d\,\xi\, (\log m)/d \log m$, where $m$ is the
mass, and $\xi\,(\log m)$ is the number of stars born per unit logarithmic
mass interval. In this parameterization, the MF for the solar neighborhood, as
derived by \cite{salpeter55} has a slope $\Gamma = -1.35$.

The MF of a system is empirically inferred by counting the number of stars in
mass intervals. As shown in previous sections, in NGC~602 the star formation
occurred over a long interval of time, and it is likely still ongoing. In these
conditions, we cannot infer the initial mass function (IMF) from the observational data without making assumptions on how the SFR varies with time. Therefore in this section we discuss only the present day mass function (PDMF) for those stars that are already visible in the optical bands, but have not evolved yet off the main sequence.

To derive the PDMF of NGC~602 we applied the same technique discussed in
\cite{sabbi07b}. This technique has the advantage of avoiding assumptions on
the age of the stellar population and therefore overcomes the age/luminosity
degeneracy which affects young star forming regions. It simply consists of
directly counting the stars between two theoretical evolutionary tracks
(Fig.~\ref{fig01}). In particular, to measure the PDMF for stars more massive
than 6 M$_\odot$ we used Padua evolutionary tracks, while we used the FRANEC
PMS tracks at lower masses.

NGC~602 is a relatively small star cluster, and its upper main sequence
appears poorly populated. As pointed out by \citet{jesus05}, in such
conditions, the size of the mass bin might affect the slope of the measured
MF, artificially increasing its steepness. To be conservative, we divided the
brightest part of the CMD into only two bins, each containing the same number
of stars.

As already done for NGC~346 \citep{sabbi07b}, we took into account the
completeness variations as a function of magnitude and color by dividing into
bins of 0.5 in magnitude and 0.25 in color, the stars identified between two
evolutionary tracks. We then normalized the number of stars lying within two
evolutionary tracks to the logarithmic width of the mass range spanned by the
tracks and to the area of the observed region, and then we applied the
appropriate completeness factor.

\begin{figure}
\plotone{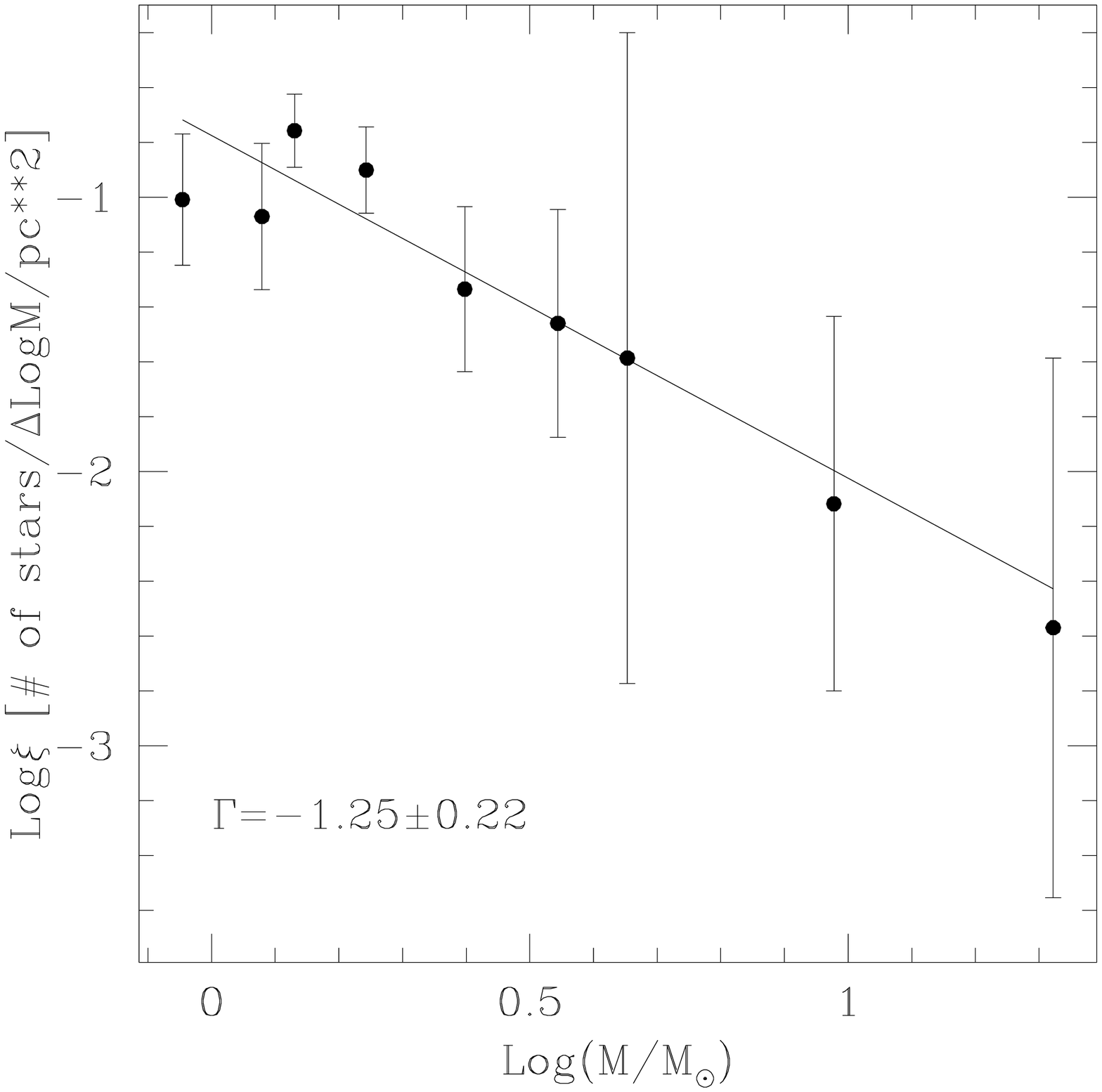}
\caption{Derived NGC~602 PDMF.}
\label{mf} 
\end{figure}

To remove the contamination of the SMC stellar population in our data, we
considered a SMC field ($\alpha = 01^h 29^m 15.9^s, \, \delta = -73\degr
42\arcmin 30\arcsec$), located at a projected distance of $\sim 47\, {\rm pc}$
from NGC~602 and observed with the F555W and F814W filters of the ACS/WFC. For
each mass bin, the estimated number of stars belonging to the field of the SMC
was subtracted from the PDMF. We therefore derived the NGC~602 PDMF between 0.7
and $30\, {\rm M}_{\odot}$ (Fig.~\ref{mf}). 
Using a weighted least mean squares fit of the data, we derive a slope
$\Gamma = -1.25 \pm 0.22$ ($\alpha=2.25$) for the PDMF of NGC~602,
in agreement with \cite{schma}. Uncertainties may still affect
the derived slope of the PDMF.  These include the percentage of
unresolved binary systems, possible variable extinction,
underestimating the mass for the most massive stars, etc. For example,
if we consider the extreme case of  a 30\% fraction of equal mass ratio  unresolved binaries, the slope of the
PDMF becomes $\Gamma=-1.48\pm 0.17$.

\subsection{The Environment}

The compelling questions we are addressing in this paper are how star
formation started and progressed in this region.  Do we observe
evidence of sequential star formation or of mass segregation?  How do
star formation and environment mutually affect each other?

It would be tempting to explain the origin of the gas shells clearly visible
in the true image (see Fig.\ref{xy}) as due to massive stellar winds and/or
supernovae explosions, but this turns out not to be the case. \cite{nigra}
used HST images and high-resolution echelle spectra from the Anglo-Australian Telescope (AAT), together  with HI survey data, to study the immediate vicinity of NGC~602 in some detail. They find that NGC~602 was probably formed $\sim 7$ Myr ago as a result of the compression and turbulence associated  with the interaction of two expanding H I shells. The AAT optical spectroscopy shows that the ionized gas close to the central cluster and within the H II region N90 is quiescent. This suggests that a supernova explosion has yet to occur within NGC~602, and that the winds from the massive stars are not sufficient to drive rapid expansion of the HII region, although they have produced a hot, X-ray emitting medium within NGC~602 (Oskinova, private communication). Nigra et al. (2008) therefore conclude that the H II region seen in Fig. 1 has been shaped by the ionizing radiation from the massive stars in the central cluster. In particular, the wealth of fine structure appears to be the result of the ionization front eroding dense, neutral, clumpy material. This picture is consistent with the idea that the ionizing radiation from the central cluster photo-evaporated the residual gas, causing star formation to cease in the central regions, while it is still ongoing in the dense, outer neutral material.

\begin{figure}
\plotone{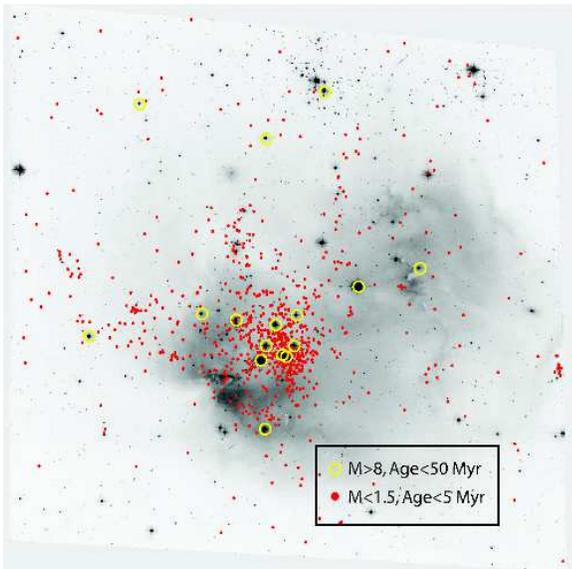}
\caption{Young stars of  masses above 8 M$_\odot$ (yellow circles) and PMS of masses below 1.5 M$_\odot$ (red dots) are superimposed on the finding chart derived from Figure \ref{xy}. Masses are deduced by isochrones fitting from Fig.~\ref{fig02}. The plotted massive stars are younger than 50 Myr and the PMS younger than 5 Myr. }
\label{figxy} 
\end{figure}

We have studied the spatial distribution of the youngest and of the most
massive stars in the region. First, we examined the young stars identified by
comparing the observed CMD with theoretical isochrones.  In Fig. \ref{figxy},
red dots indicate PMS stars, younger than 5 Myr, with masses between 0.5 and
1.5 $M_{\odot}$. Most of these stars are concentrated in the core of the
central cluster and along several filamentary structures. Interestingly, this
population is almost absent in the two sub-clusters NGC~602-B and NGC~602-B2. A
first obvious conclusion is that NGC~602-B and NGC~602-B2 are slightly older
than NGC~602-A. As already mentioned above, we do not have enough stars to infer if NGC~602-A, B and B2 are or not coeval, however it is clear that the three regions had significantly different SFRs. 

Figure \ref{figxy} shows also the position (yellow large circles) of
stars more massive than 8 $M_{\odot}$ in any evolutionary phase.
These stars have formed in the last 50 Myr or so.  They are mostly
concentrated in the center, but several are scattered outside the central
cluster, we can therefore conclude that we are not seeing evidence of mass
segregation in this cluster. Given the relatively young age of the central
cluster, it is also appropriate to infer that the massive stars are formed in
situ, since the cluster has not had enough time to undergo any dynamical
evolution.

To understand the progression of star formation in the
region, it is also fundamental to note  the  result
of \citet{carlson07}, who found, from Spitzer data, a generation of class 0.5
- I YSO (age 10$^{4}$ - 10$^{5}$ yr) located at the outskirts of the central
cluster, along the dusty ridges. Carlson et al. (2007) concluded that star
formation started at the center of the cluster and likely propagated outwards,
with star formation still ongoing in the outer dusty regions. 

\begin{figure*}
\centering \includegraphics[width=10cm]{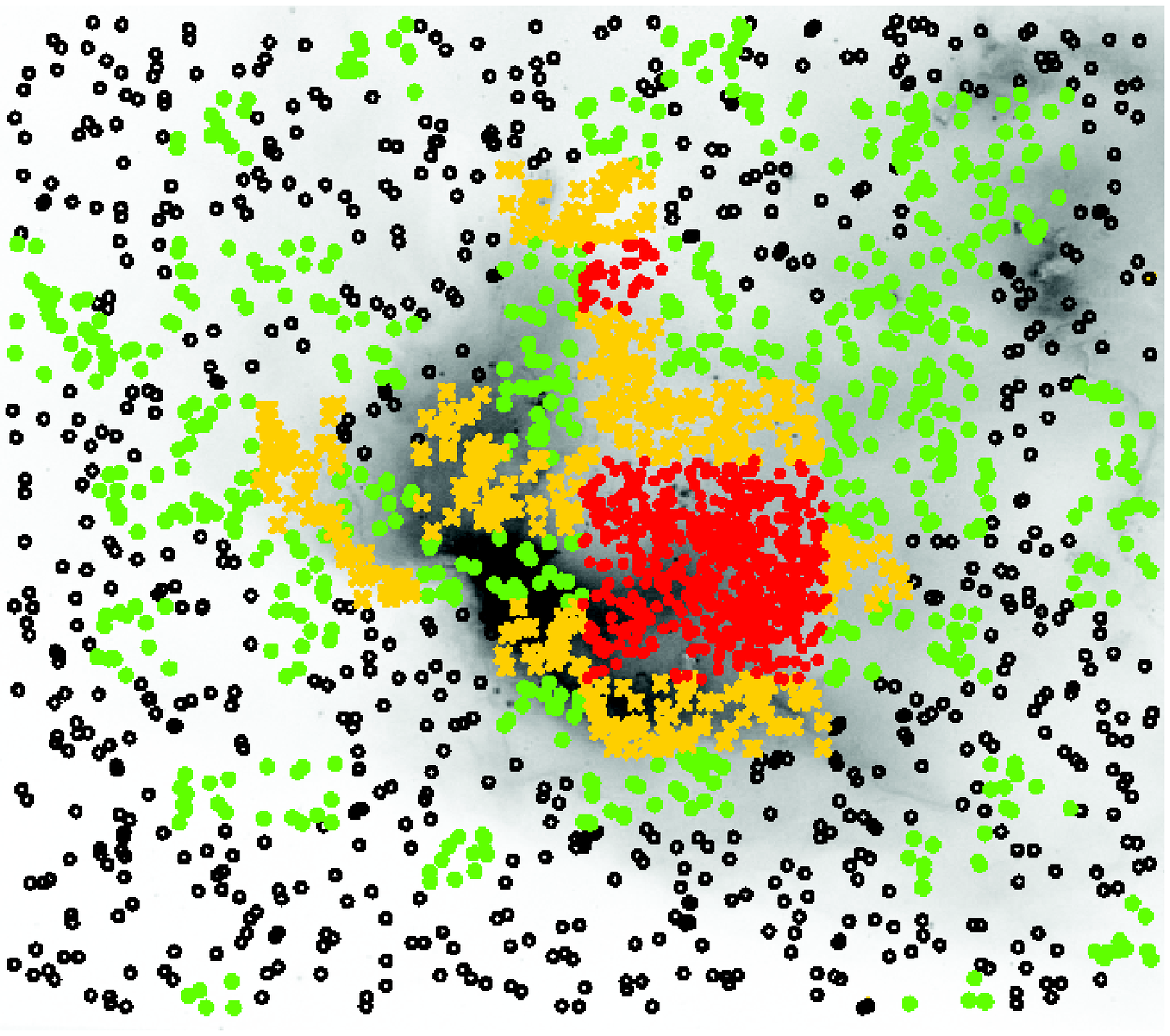}\\
\centering \includegraphics[width=6cm]{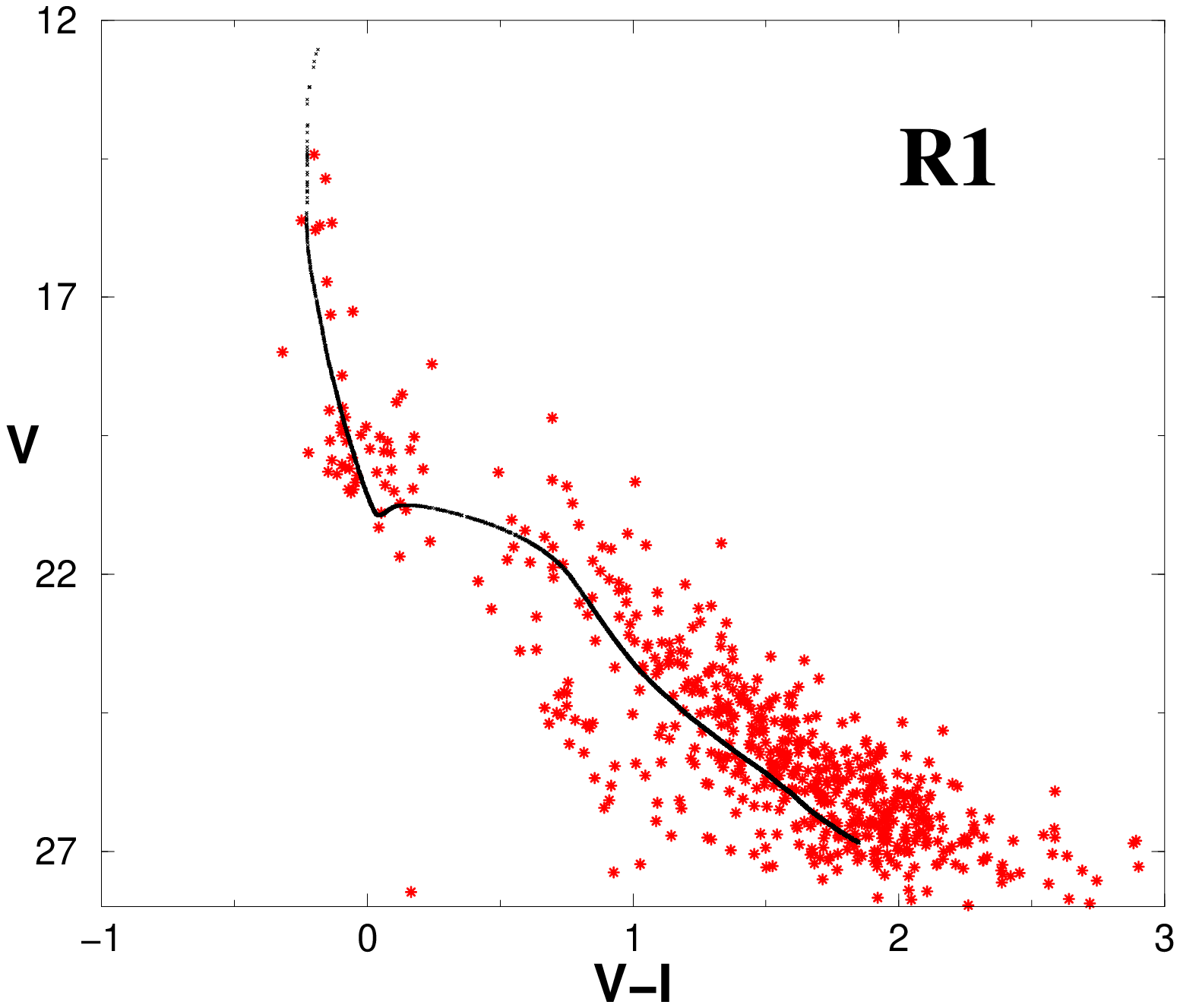}
\centering \includegraphics[width=6cm]{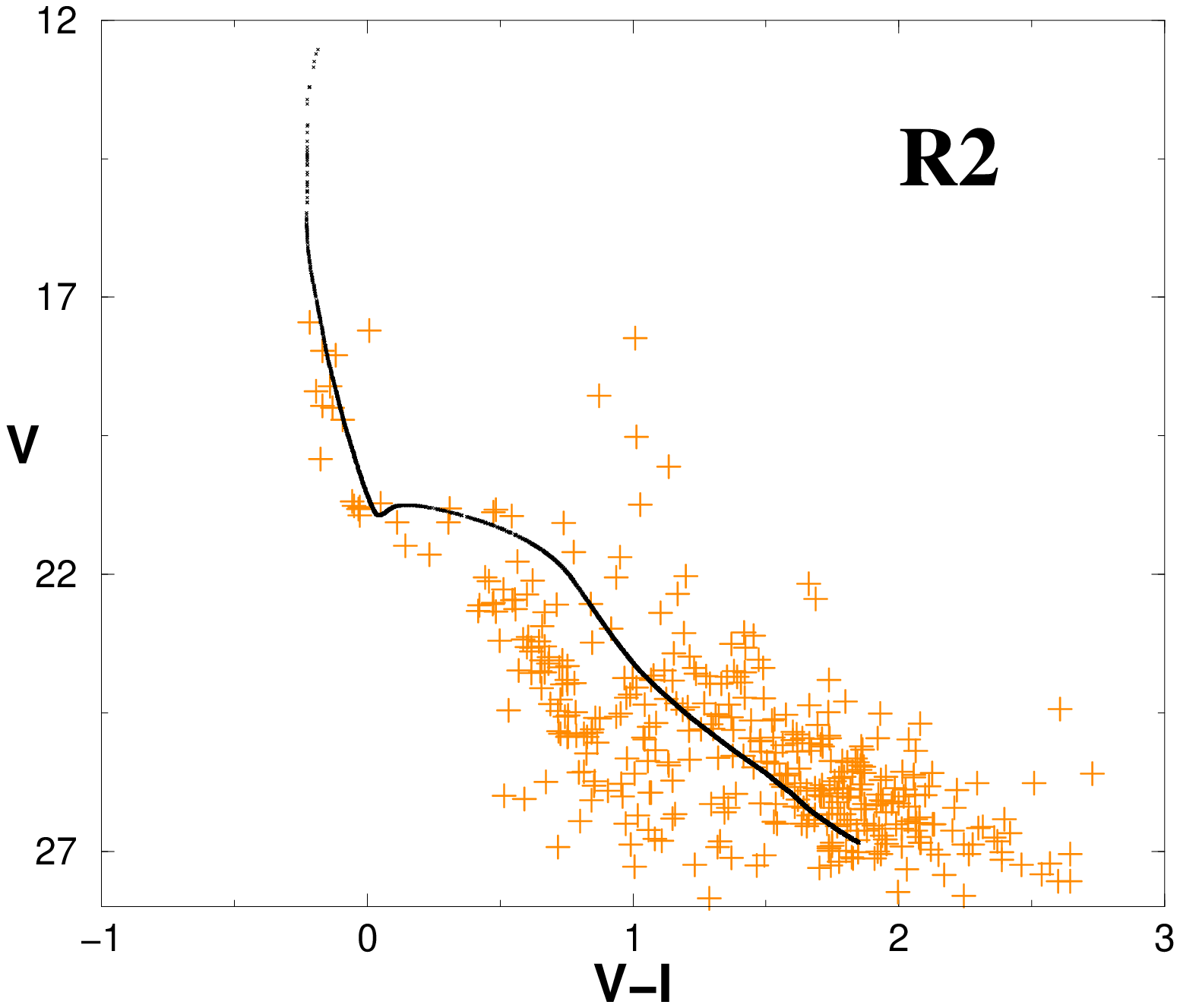}\\
\centering \includegraphics[width=6cm]{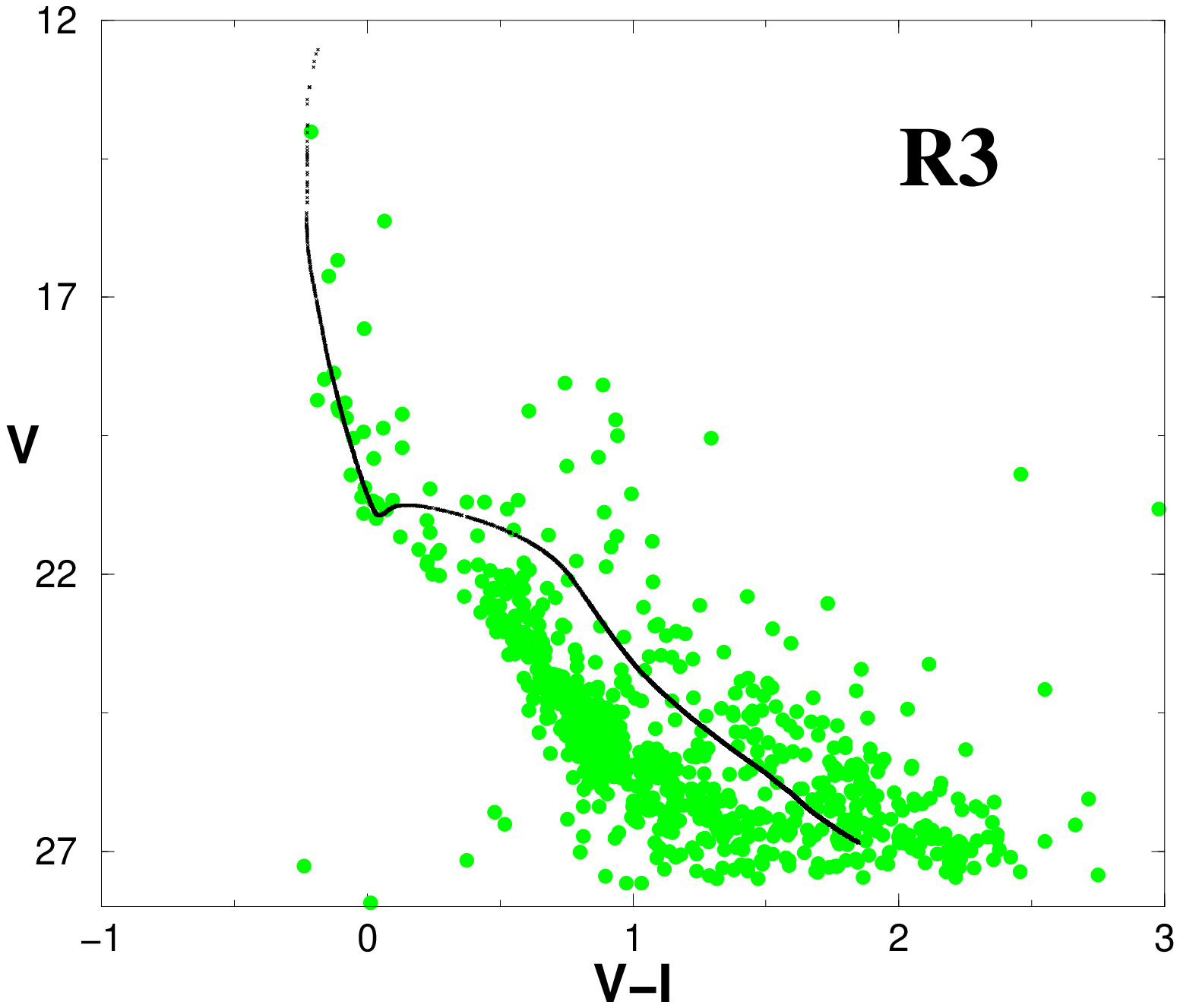}
\centering \includegraphics[width=6cm]{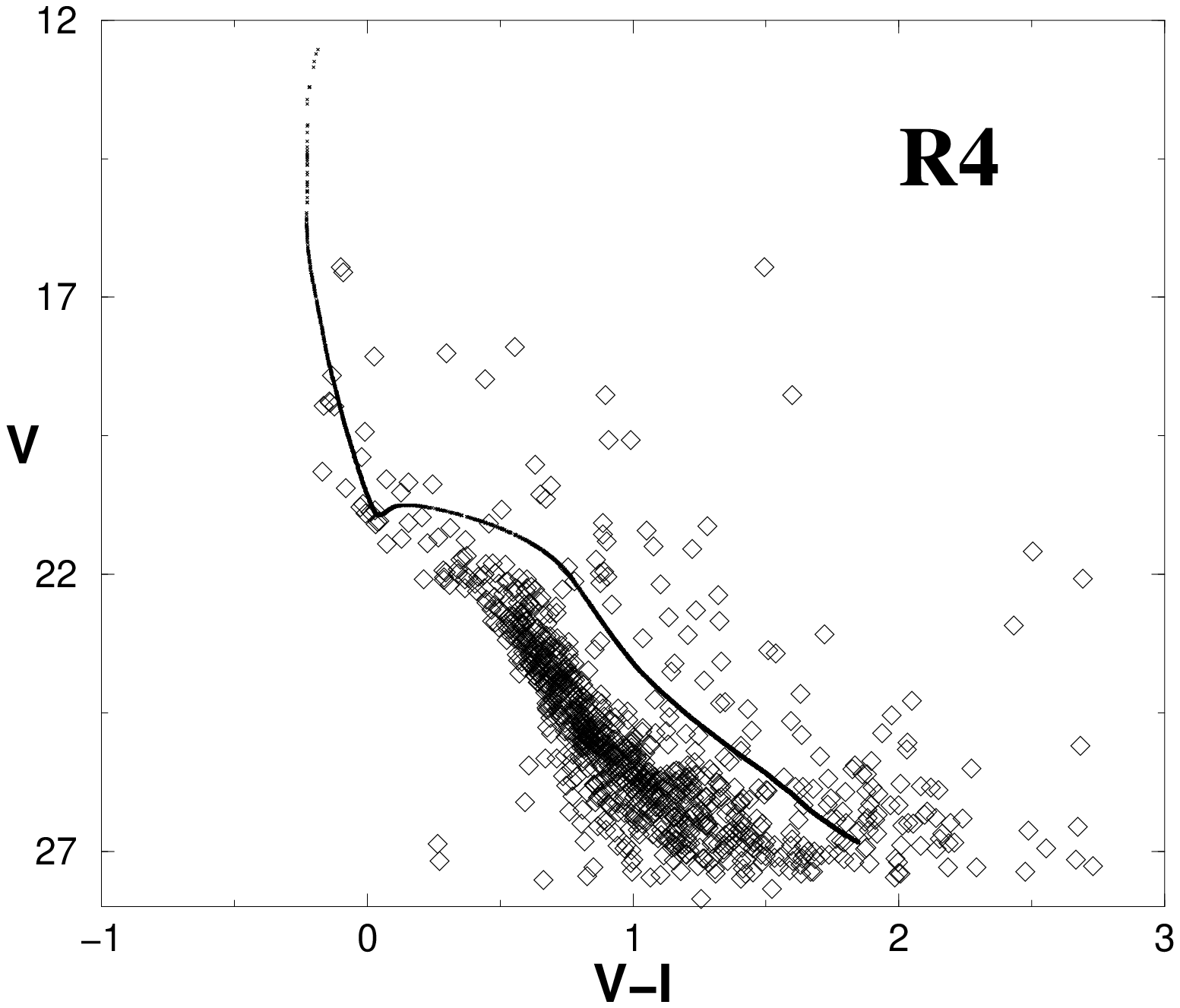}
\caption{Top panel: XY map of NGC~602-A and its neighborhood. The different
colors represent different constant density surfaces (see text). Lower panels
show the separate CMDs of the four density regions. From R1 to R4
the stellar crowding decreases.}
\label{figxy2} 
\end{figure*}

In order to investigate the relation, if any, of the recent star
formation activity with the current stellar density we have divided
the region around NGC~602-A in bins of $200\times200$ pixels
($10\arcsec \times 10\arcsec$) and counted the number of stars in each
bin.  The bins have then been grouped according to their number of
stars: bins with more than 30 stars are defined as R1, those with
20-30 stars as R2, those with 10-20 stars as R3, and those with less
than 10 stars as R4. The spatial distribution of the four density
intervals is shown in the top panel of Figure \ref{figxy2}, where
individual stars are plotted with different symbols. Stars in the
highest density bins R1 are represented by red dots, stars in R2 by
orange asterisks, stars in R3 by green dots, and stars in R4 by black
open circles. This map suggests the presence of a high density region
concentrated on NGC~602-A, surrounded by filamentary structures of
decreasing stellar density and super -imposed on a rather homogeneous
field, with minimum density, possibly corresponding to the SMC
background.

The CMDs corresponding to each of the four density intervals are
presented in the other panels of Fig. \ref{figxy2}. The 5 Myr
isochrone is also plotted for reference in each panel.  These CMDs
clearly show that the high density regions R1 and R2 are dominated by
PMS stars, whose distribution is tight and close to the 5 Myr
isochrone, whilst very few PMS stars are detected in the lowest
density region R4 (black dots). If we consider the number of PMS per
unit area, we find that the PMS density drops by a factor of 70-80
from R1 to R4 (the corresponding density ratios from R1 to R2 and R3
being respectively 3 and 14). At first glance, the lower main sequence
appears instead more populated moving from high to low-density
regions. However, once the star counts are normalized to the area, we
find that the main sequence density drops too, although only by a
factor 1.5-2.  While the PMS stars show a clear trend of decreasing density
with increasing distance from the center of NGC~602-A, the MS only
shows a discontinuity between the external, low density region R4 and
all the others, which have MS densities similar to each other.  We
interpret the moderate discontinuity in MS density from R4 to the
denser regions as due to the simultaneous contribution of field plus
cluster MSs in R2-3-4.

The clear star density gradient indicates that the recent SF rate has been
higher in the central regions. However, when we take into account that the 25
{\it bona fide} YSOs identified by \citet{carlson07} are found on the gas rims
and are absent in the more central regions, we need to conclude that the most
recent SF activity has occurred at the outskirts of the cluster. This suggests
that the SF activity has stopped earlier in the center than in the periphery
of the cluster, probably because of the higher gas consumption rate.

\section{Synthetic CMDs: the approach}

Synthetic stellar populations are generated to reproduce the observed CMD. The
synthetic CMD is created following the procedure originally described by
\cite{tosi} and subsequently updated \citep{greggio, ange05}: 1) stellar
masses and ages are randomly chosen from a time independent Initial Mass
Function and a star formation law, respectively; 2) an assumed age-metallicity
relation provides the metallicity; 3) stellar tracks ranging from pre-main
sequence to asymptotic giant branch phase are interpolated, deriving the
absolute photometry for the synthetic population; 4) distance, reddening,
incompleteness and photometric errors are applied. Once a synthetic CMD is
created, we apply a maximum likelihood classification algorithm as presented
in \cite{cigno06} in order to make a rigorous comparison with the
observations. To evaluate the goodness of the assumed model, observational and
theoretical CMDs are converted into two dimensional histograms by dividing the
CMDs in cells ($0.1\times 0.1$ mags) and counting the stars in each cell. Once
the number of theoretical and observational objects is known in each cell, we
implement a function of the residuals (Poissonian ${\chi}^2$) to quantify the
differences between the two histograms. This function is minimized by means of
a downhill simplex algorithm. In order to avoid local minima, the simplex is
re-started from thousands of initial random positions. A bootstrap method is
used to assess a confidence around the best parameters. The entire process is
repeated for each bootstrapped data set. The error bars represent the mean
deviation using 100 bootstrap. For a detailed description of this procedure
see \cite{cigno06}.

\begin{figure}
\plotone{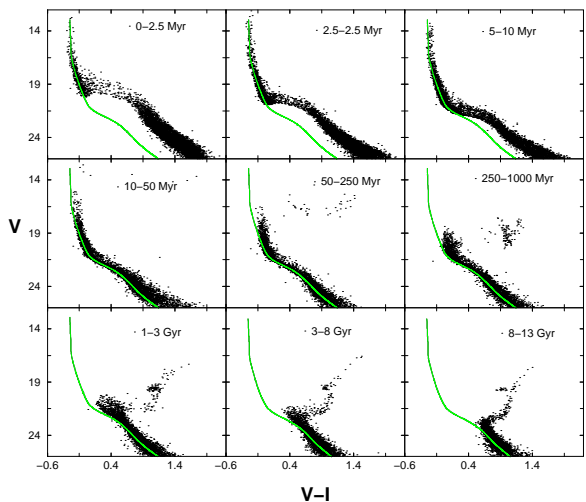}
\caption{\small Synthetic partial CMDs for the labeled time interval,
 with 30\% of unresolved binaries being assumed. The continuous line
 is the zero age main sequence (ZAMS). Upper three panels: the star
 formation activity is younger than 10 Myr and consequently most of
 the stars are still in PMS (only massive stars are in main
 sequence). As expected, aging the system (moving from the left panel
 to the right panel) shifts the \emph{turn-on} point (where the PMS
 stars reach the main sequence) to progressively fainter
 magnitudes. Middle three panels: most of the stars have reached the
 main sequence and some massive stars have evolved off (core helium
 burning stars). Lower panels: the oldest star formation steps.  Here
 also intermediate mass stars (roughly $M>1 M_{\odot}$) have evolved
 off the main sequence.}
\label{cmds} 
\end{figure}

In order to reduce the computational time, the star formation rate is
parameterized as a set of constant values over adjacent temporal
intervals. Each constant star formation produces a partial CMD (a quas-single
stellar population) and a generic star formation history can be written as a
linear combination of the partial CMDs. The advantage of a similar procedure
is that a library of partial CMDs is built (see Figure \ref{cmds}) just once,
in advance, and the genesis of any additional SFH is reduced to elementary
algebraic combinations. Finding the best SFH means searching for the
combination of partial CMDs which minimizes the $\chi^2$.

Disentangling a stellar population showing both very recent (Myr) and very old
(Gyr) episodes of star formation is not straightforward. Only low mass stars will survive
from ancient episodes because their evolutionary timescales are very long:
small CMD displacements, for example due to photometric errors, can bias the
age estimates up to some Gyr. In this case, increasing the time resolution,
besides being time-consuming, may produce unrealistic star formation rates due
to misinterpretation. Hence, the choice of temporal resolution must follow
both the time scale of the underlying stellar populations and the data scatter
(photometric errors, incompleteness, etc.). Given the overall recent activity
of our field, a practical way out is to use a coarser temporal resolution for
the older epochs, which: 1) allows us to avoid SFH artifacts at early epochs;
2) reduces the Poisson noise; 3) reduces the parameter space. In our
simulations the duration of each step is progressively increased with age as
labeled in Figure \ref{cmds}.

\section{Recovering the star formation history}
\begin{figure}
\plotone{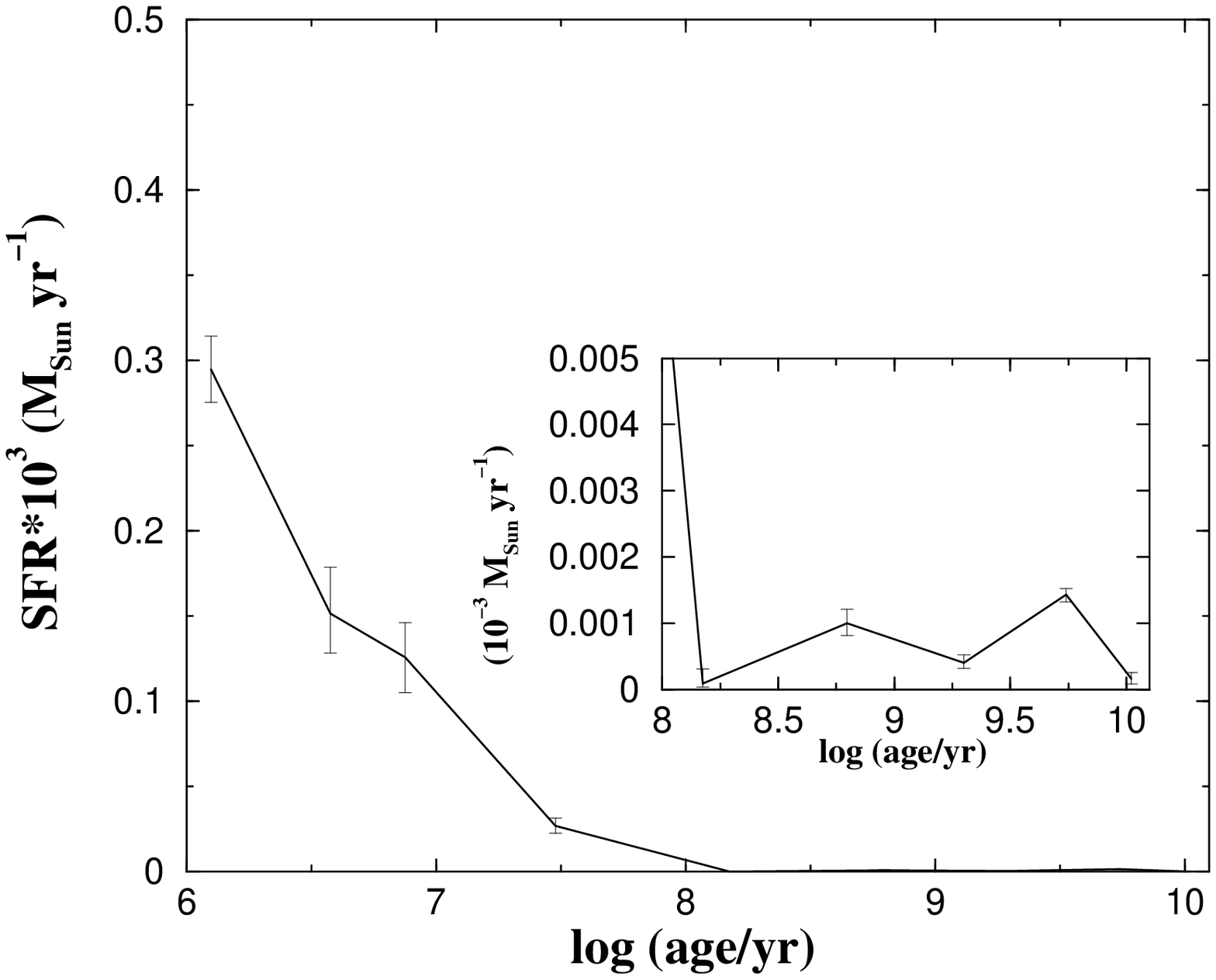}
\caption{Recovered star formation rate using an IMF exponent $\alpha=2.5$ and a
binary fraction of 30\%. The old star formation activity is enlarged in the
small panel.}
\label{figsfr} 
\end{figure}

Following the prescriptions of Section 5 and the population
properties described in Section 4 we assume that objects younger
than 3 Gyr are formed with metallicity $Z=0.004$, and that the older
ones are formed with $Z=0.001$. For all synthetic stars, a distance
modulus ${(m-M)_0}=18.9$ and reddening $E(B-V)\sim 0.08$ are
assumed. Before exploring different IMF exponents and binary
recipes, the SFH is searched starting from an arbitrary power-law IMF ($\phi (m)\propto m^{-\alpha}$) and binary fraction and then letting them vary. At the end of the iterations, the best SFH turned out to have $\alpha$=2.5 and a binary fraction of 30\% (companion masses are picked randomly from the
same IMF). Our result is shown in Figure \ref{figsfr}.
The maximum activity is reached in the last 2.5 Myr (about $0.3\times
10^{-3}\,\,M_{\odot}\,yr^{-1}$). This phenomenon is also confirmed through a
visual inspection of the CMD morphology. The PMS phase is well separated from
the main sequence, suggesting that most of the PMS stars are moving down their
own Hayashi tracks. Looking back in time, the star forming activity drops by a
factor approximately 2 between 2.5 and 5 Myr ago and a factor of 10 between 10
Myr and 50 Myr ago. Beyond this epoch, it drops again by another factor of 10,
but slightly increasing around 6-8 Gyr ago. Finally, the period between 8 and
13 Gyr ago is characterized by a very low activity, confirming that the bulk
of the oldest stars in our field have ages of about 8 Gyr.

In spite of the fact that the stronger star formation activity is extremely
recent, most of the stellar mass budget (the total mass of all stars ever born
in our field, i.e., the integral of the star formation history) was forged
into stars at earlier epochs. Considering only stars more massive than
$0.45\,M_{\odot}$, the NGC~602 field has produced about $4400\,M_{\odot}$ in
the last 3 Gyr, while as much as $8000\,M_{\odot}$ were converted into stars
between 3 and 13 Gyr ($7000\,M_{\odot}$ between 3 and 8 Gyr ago). Looking at
Figure \ref{figxy} we see that stars younger than 5 Myr are confined in a
surface of roughly $10\times 10\,pc^2$: considering that the mean SFR in the
last 5 Myr is about $0.22\times 10^{-3}\,\,M_{\odot}\,yr^{-1}$, it implies a
recent star formation rate density of $2.2\times
10^{-6}\,\,M_{\odot}\,yr^{-1}\,pc^{-2}$ (i.e. $2.2\times
\,M_{\odot}\,yr^{-1}\,kpc^{-2}$). Extrapolating the IMF down to the hydrogen
burning limit ($\sim 0.1\,M_{\odot}$) with the same slope, this number
increases to $4.8 \times 10^{-6}\,\,M_{\odot}\,yr^{-1}\,pc^{-2}$, while no
significant variations exist if the IMF flattens below
$0.45\,M_{\odot}$ (\citealt{k01} finds $\alpha=1.3\pm\,0.5$).

\begin{figure}
\centering \includegraphics[width=7cm]{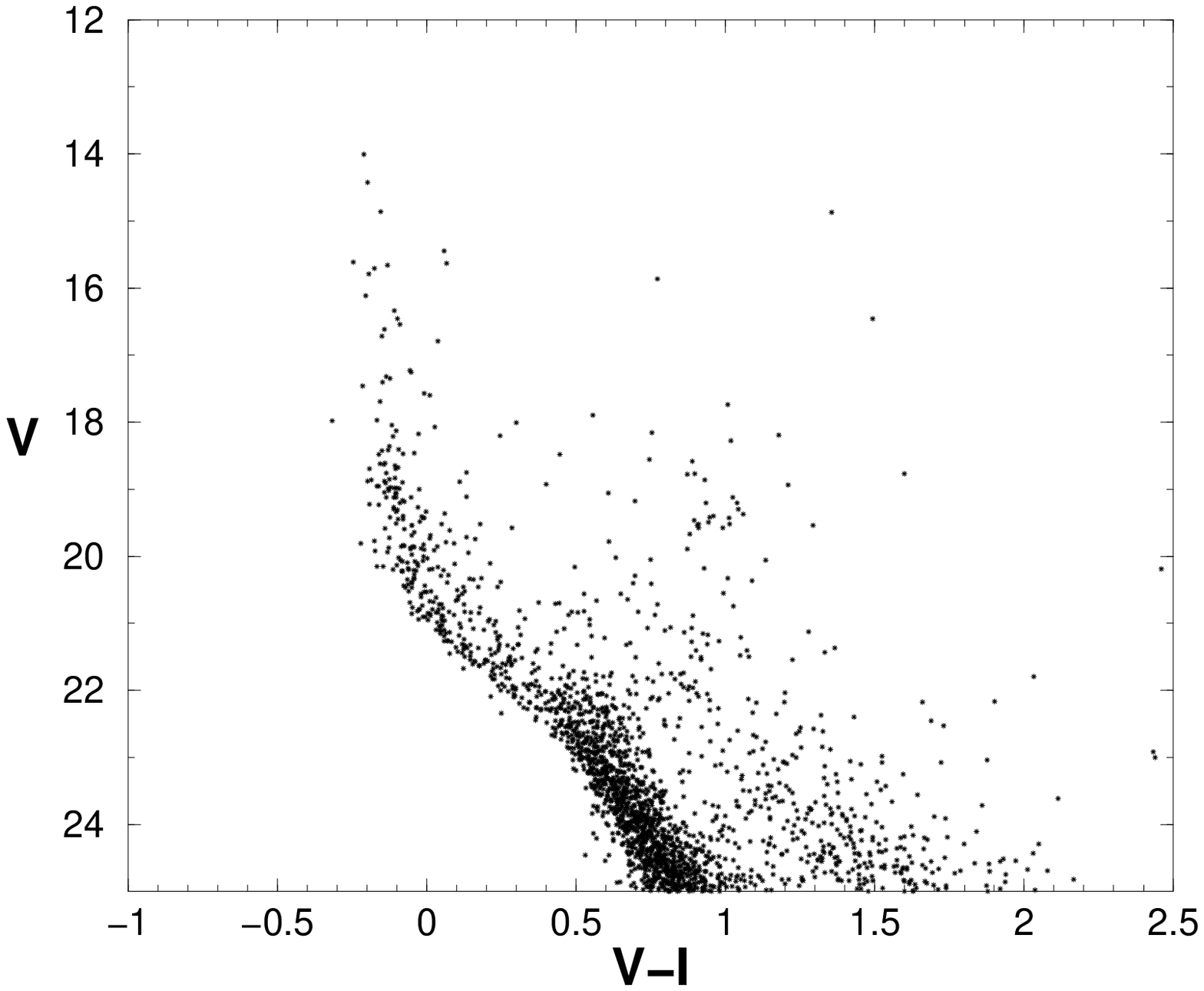}\\
\centering \includegraphics[width=7cm]{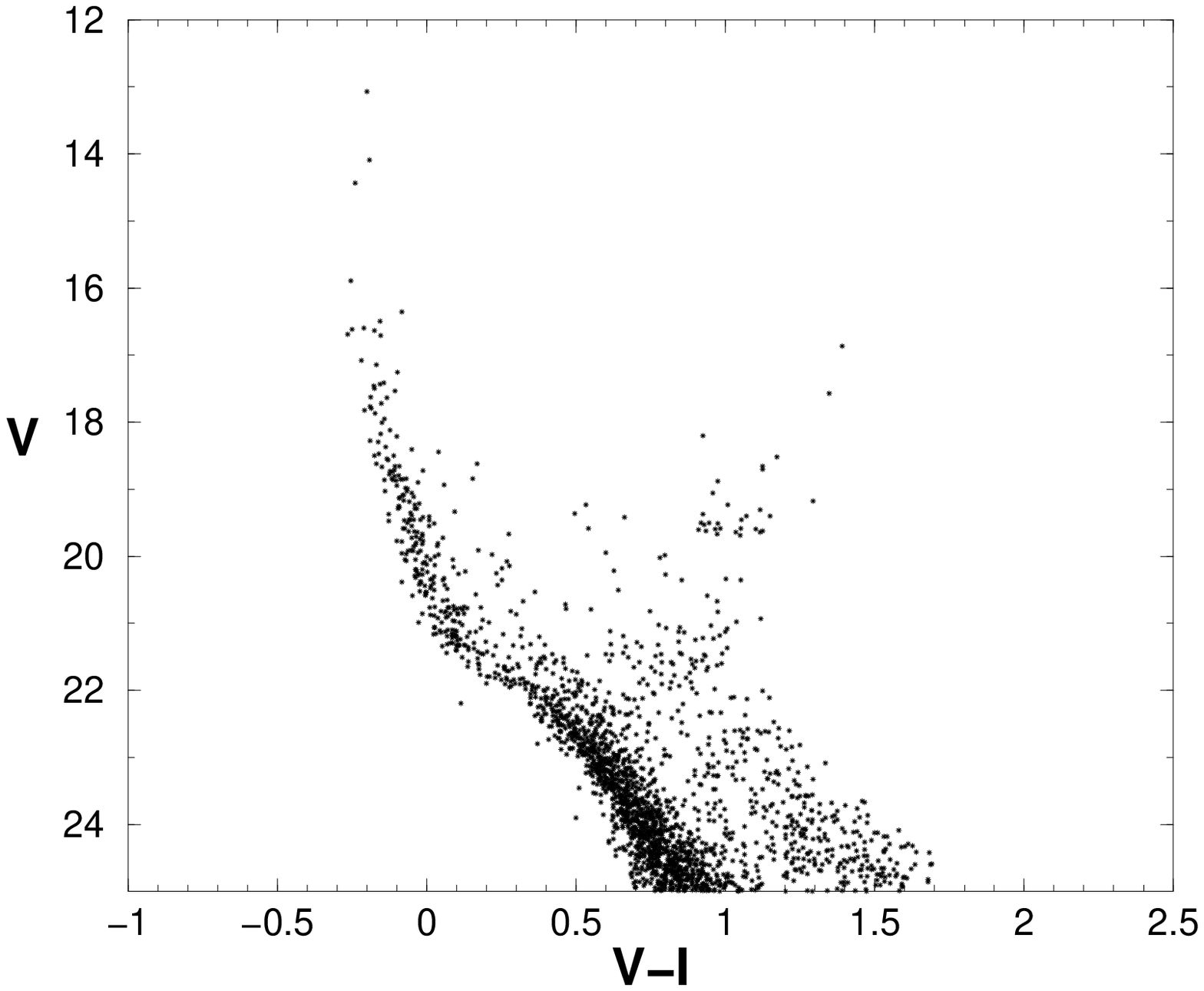}
\caption{Up: Observational CMD. Bottom: Best synthetic CMD. It is obtained
assuming an IMF exponent $\alpha=2.5$, a binary fraction of 30\% and the SFH
of Fig. \ref{figsfr}.}
\label{synth} 
\end{figure}

The CMD corresponding to our best SFH is shown in the bottom panel of
Fig.\ref{synth}. The top panel of the same Figure shows the observed CMD for a
direct comparison.

\subsection{The robustness of the SFH}

In order to check the robustness of the resulting SFH, it is important
to carry out a stability analysis. Indeed, our result has been
obtained under specific assumptions on secondary parameters (IMF,
binary fraction, distance and reddening), which are intrinsically
poorly known. We dedicate the next sections to quantifying the
influence on the recovered SFH by these uncertiainties. The distance
modulus of the young population is varied in the range 18.7-18.9. For
the IMF we have tested $\alpha$ values between 2 and 3 (Salpeter's
being 2.35); for the binary stars we have assumed fractions from 20 \%
to 70 \%.

\subsubsection{Distance}
\begin{figure}
\plotone{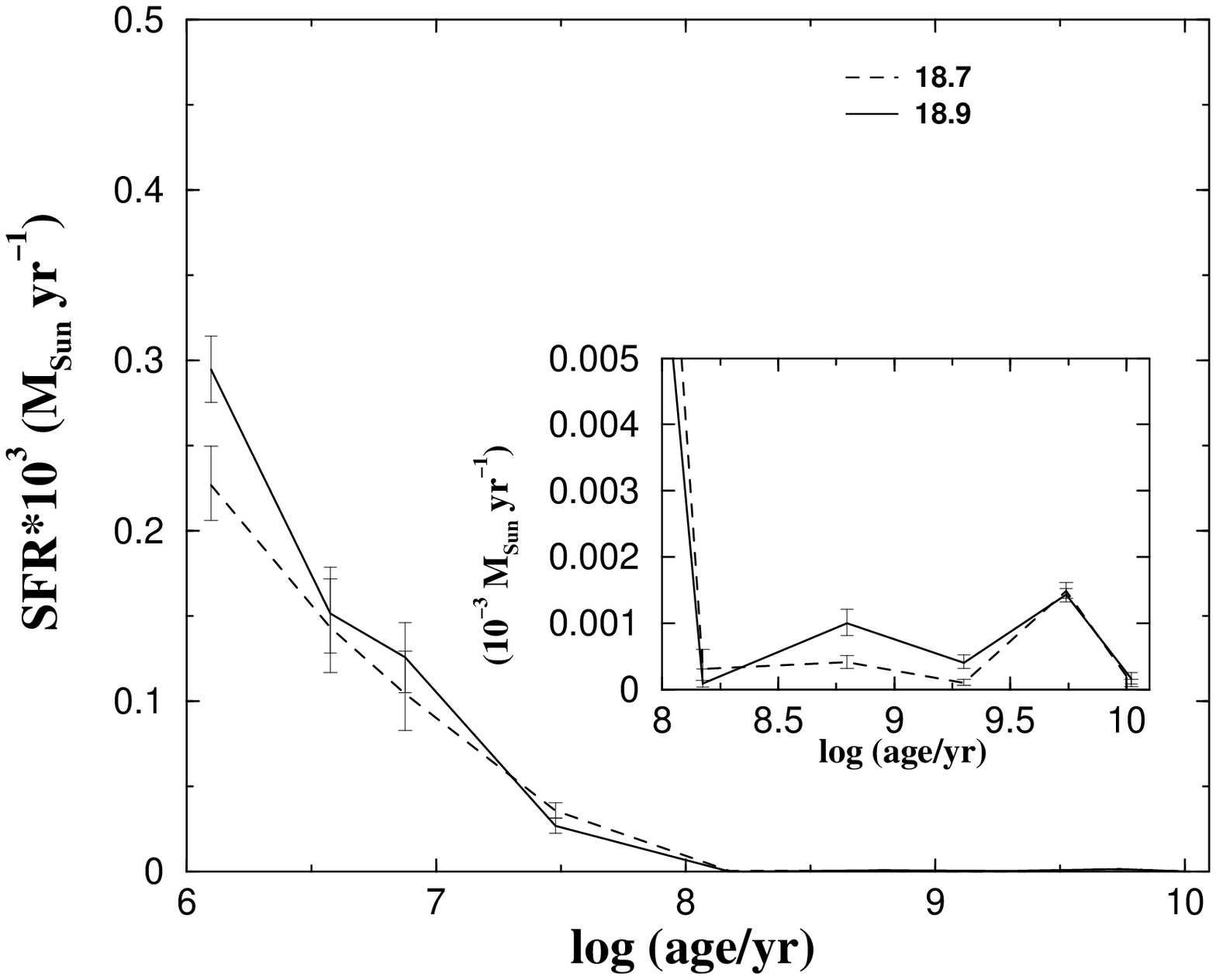}
\caption{Recovered SFH for two choices of the young population distance
modulus.}
\label{sfrdist} 
\end{figure}

While the luminosity of the red clump allows us to tightly constrain the distance
modulus of the old population at 18.9, the distance of the NGC~602 young cluster
remains more uncertain because of the MS spread. A distance modulus of 18.7
would better match the location of the young population in the CMD.  This
value would place the cluster slightly in front of the SMC, confirming that this
object is a member of the "wing". In Fig.\ref{sfrdist} we show together the
two SFHs derived for the two adopted values of the distance modulus, namely
18.7 and 18.9.  It is evident that the SFH is only weakly affected by the
slight change in this parameter.

\subsubsection{IMF}
\begin{figure}
\plotone{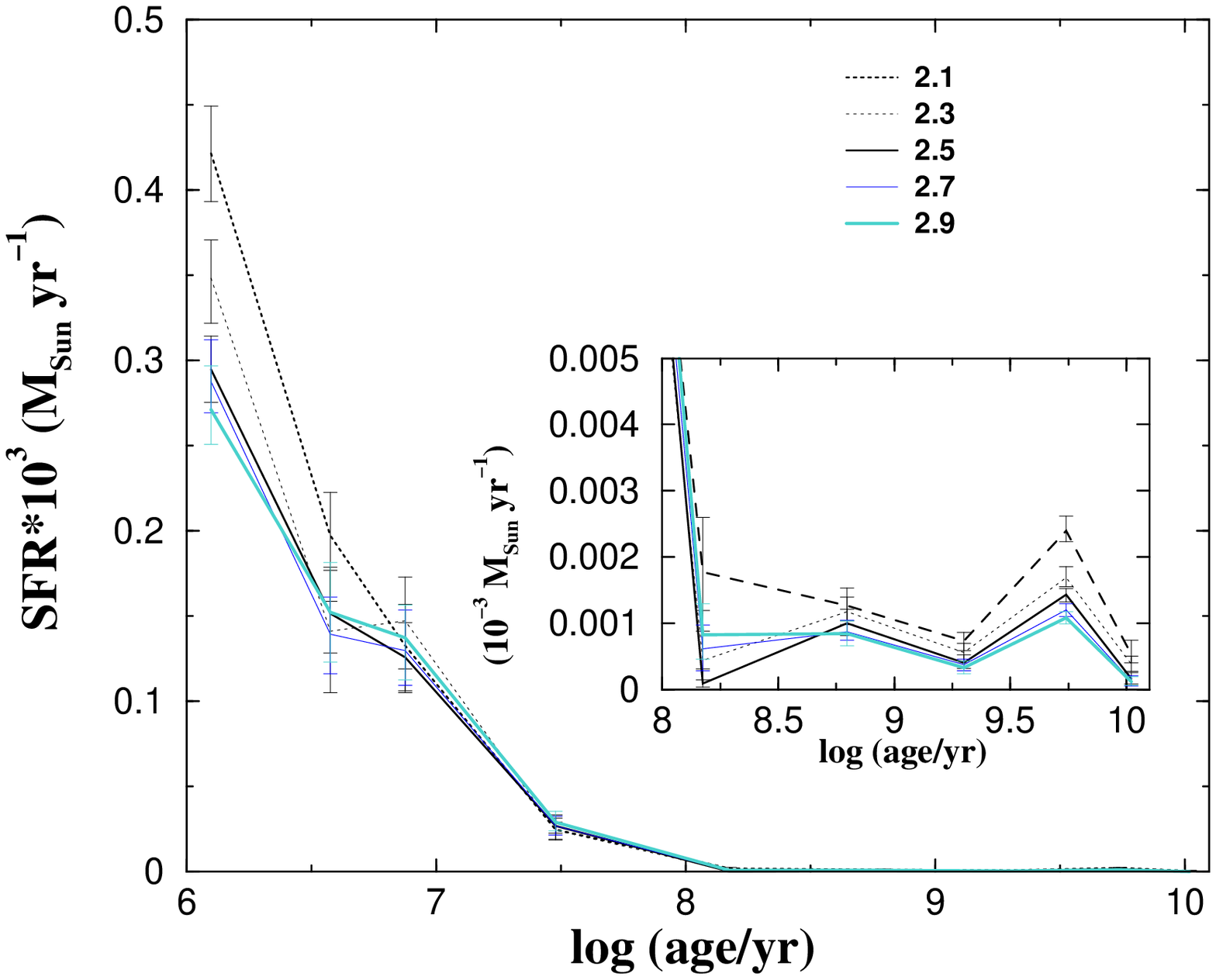}
\caption{Recovered SFH for different assumptions on the IMF exponent.}
\label{imf} 
\end{figure}

To assess how the recovered SFH is influenced by the IMF, the star formation
is re-determined adopting different IMF exponents, assuming in this case a
fixed binary population of 30\%. Figure \ref{imf} shows that the mean SFH
systematically increases with steepening IMF. In fact, for a fixed total mass,
a shallower IMF implies fewer stars surviving in the final sample (because the
IMF flattening causes a deficit of low-mass long-lived stars), hence a higher
SFH is required to produce the observed star counts.

The overall trend of the SF activity with time does not change significantly,
though, varying at most by a factor smaller than
2. We can thus conclude that the recovered SFH is quite robust against IMF
variations.

\begin{figure}
\plotone{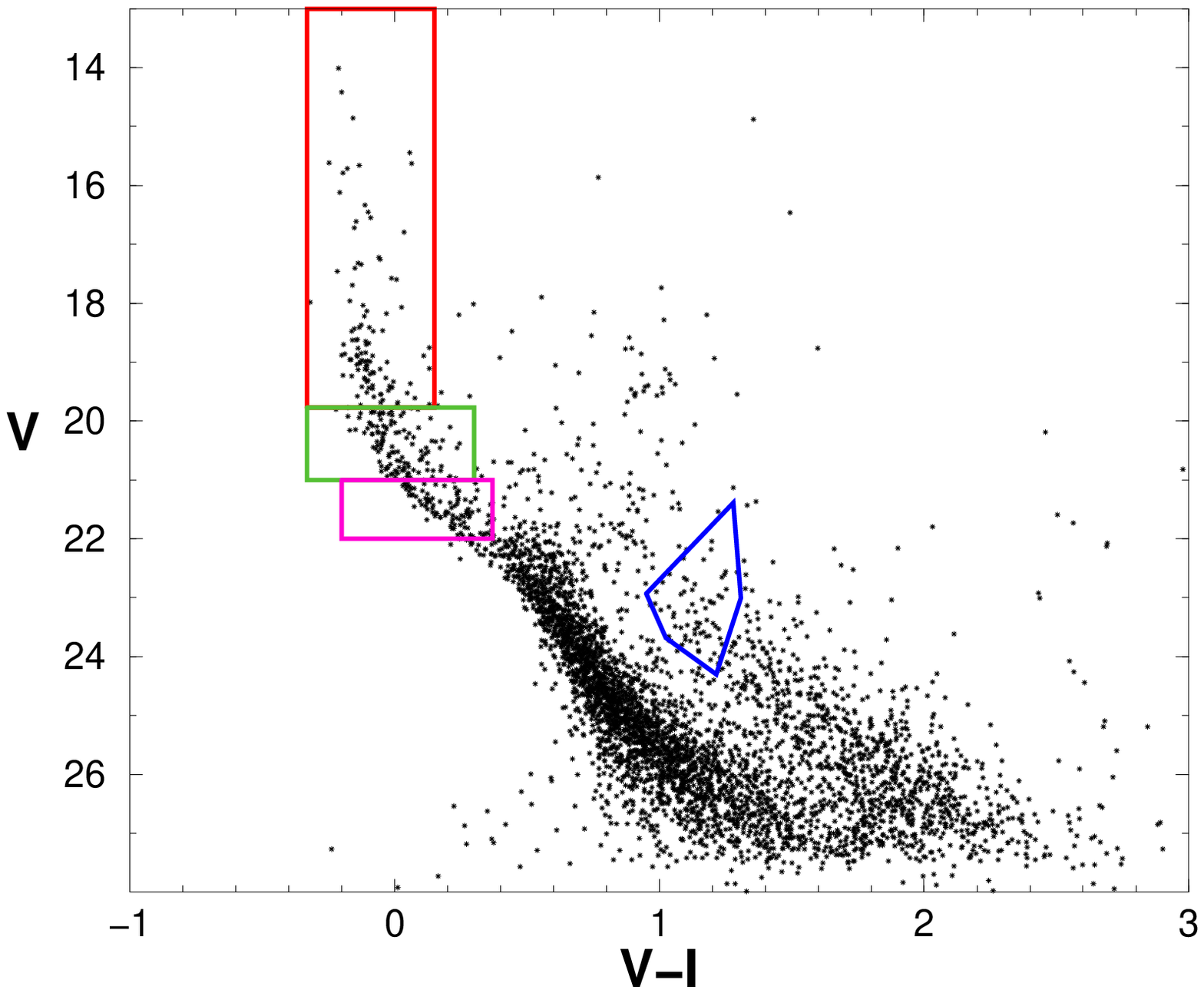}
\caption{CMD regions used to constrain the IMF (see text).}
\label{box} 
\end{figure}

We also tried to further constrain the IMF exponent. However, instead of
adding an extra free parameter (the IMF exponent) in the whole procedure to
infer the SFH, only the best SFHs of Figure \ref{imf} has been explored
further. In particular, we tested which IMF exponent produces, within the
Poisson noise, acceptable star counts in relatively large color-magnitude
boxes (see Figure \ref{box}).  We map those stellar phases which are
especially sensitive to the IMF (for example, the ratio between the star
counts from the upper main sequence stars and the PMS is extremely sensitive to
the IMF slope).

With respect to the likelihood method, this strategy allows a strong
improvement in statistics. As a further check, the luminosity function for
synthetic stars brighter than $V=24$ has been compared with the observational
luminosity function.

\begin{figure}
\centering \includegraphics[width=6cm]{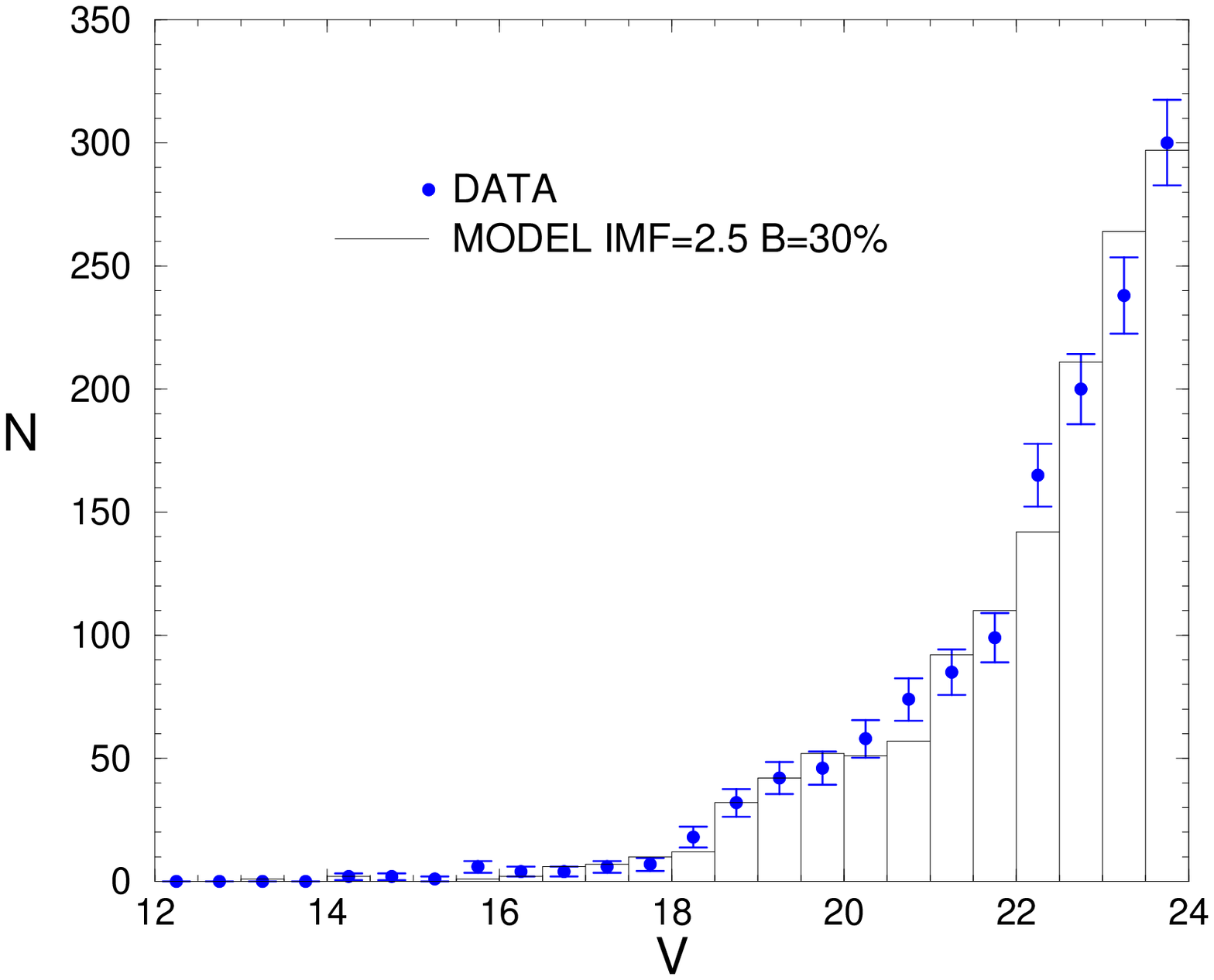}\\
\centering \includegraphics[width=6cm]{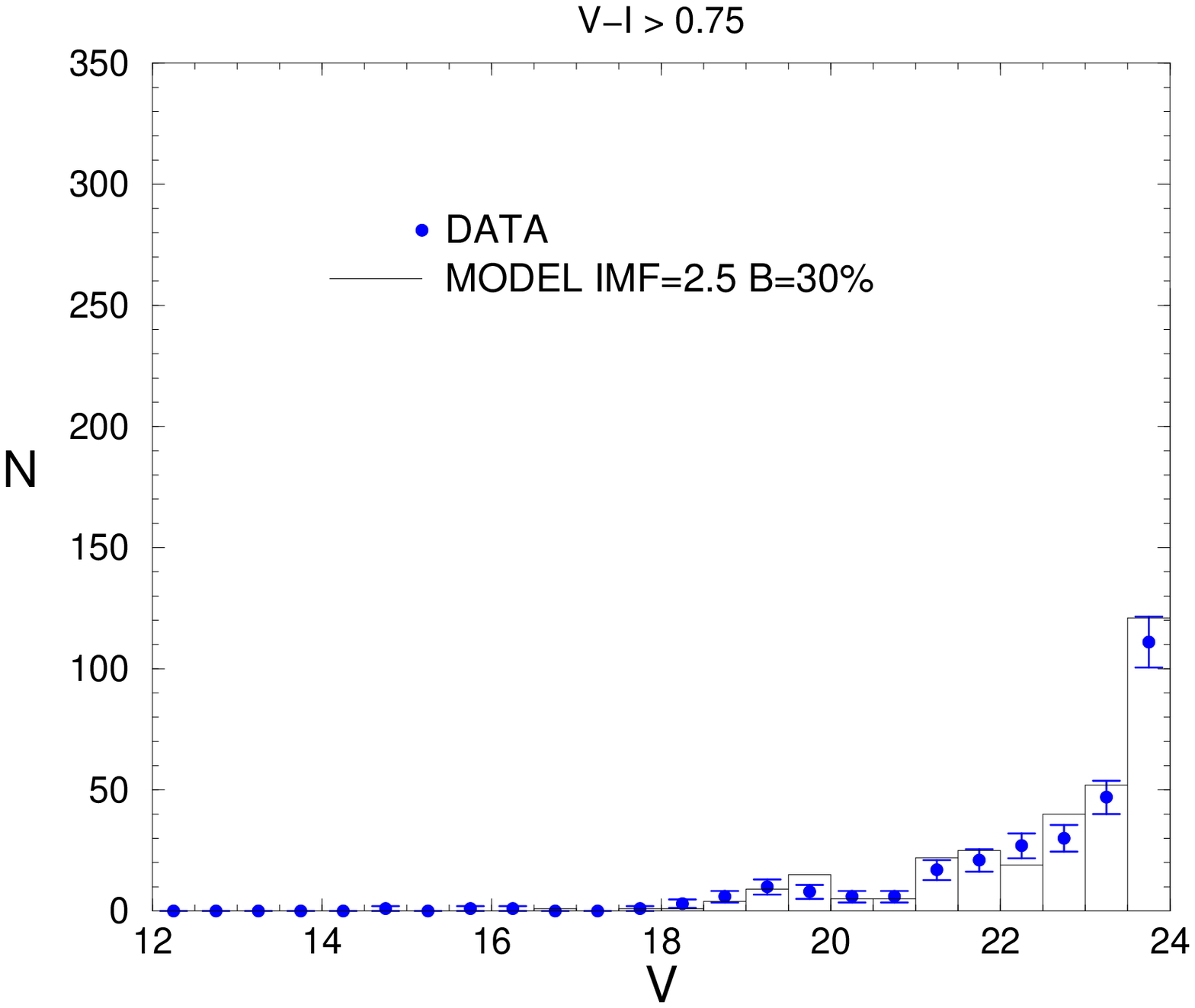}\\
\centering \includegraphics[width=6cm]{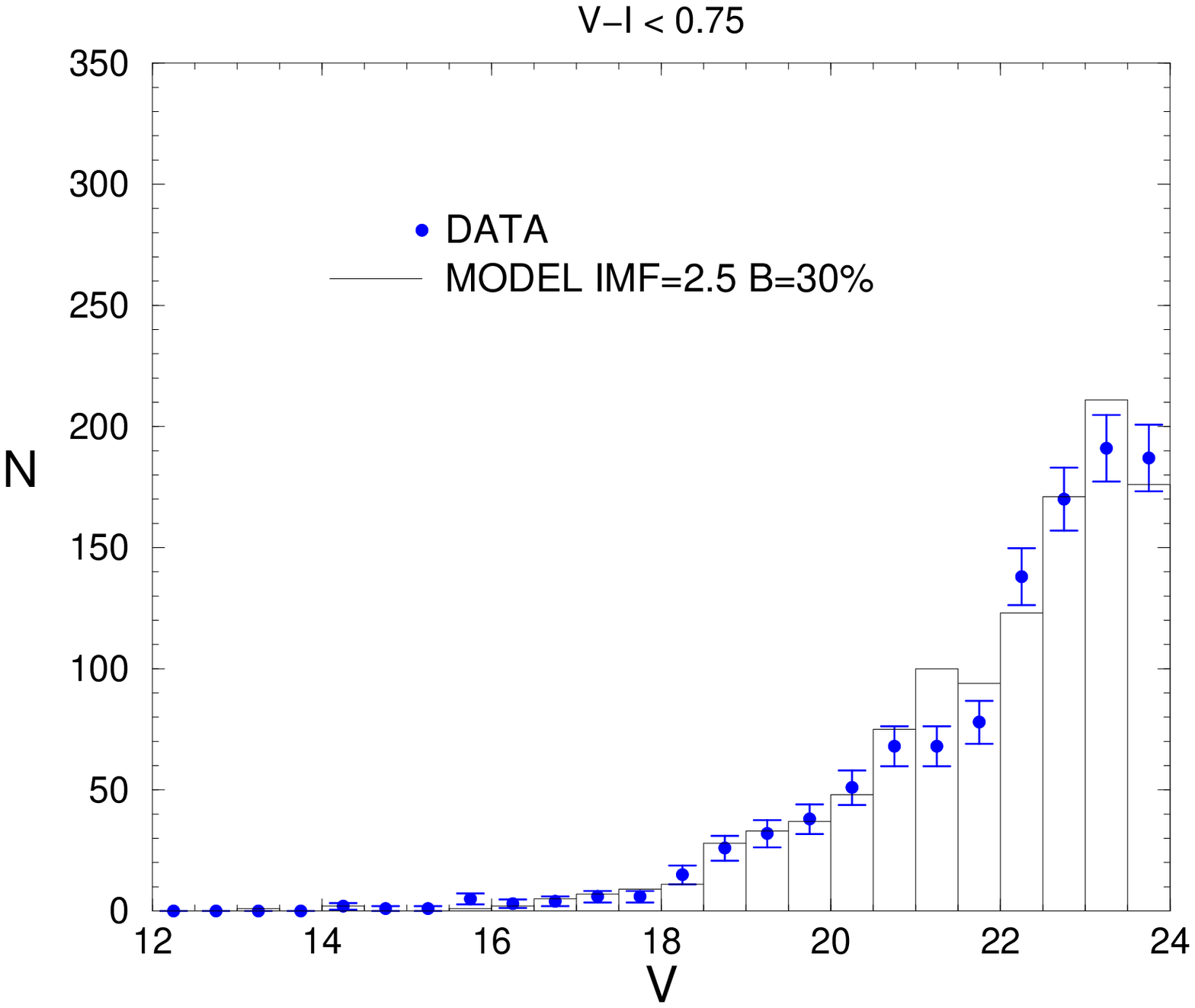}
\caption{Simulated (IMF exponent $\alpha=2.5$ and binary fraction 30\%) versus
observational luminosity function for the whole sample (top panel), stars
with $V-I>0.75$ (middle panel), stars with $V-I<0.75$ (bottom panel).} Unfortunately, the small size of our sample does not allow us to put tighter constraints.

\label{lfs1} 
\end{figure}

\begin{figure}
\centering \includegraphics[width=6cm]{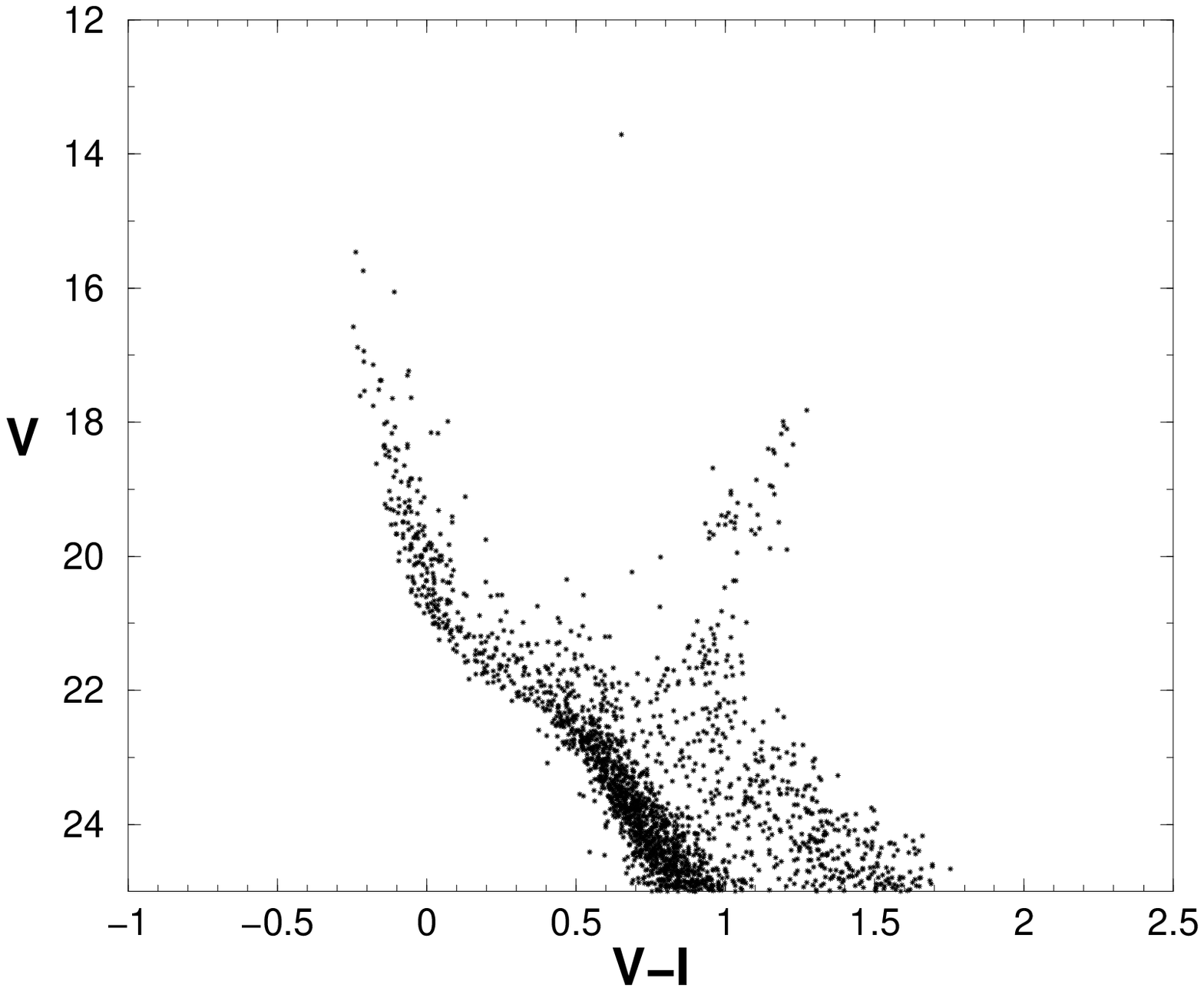}\\
\centering \includegraphics[width=6cm]{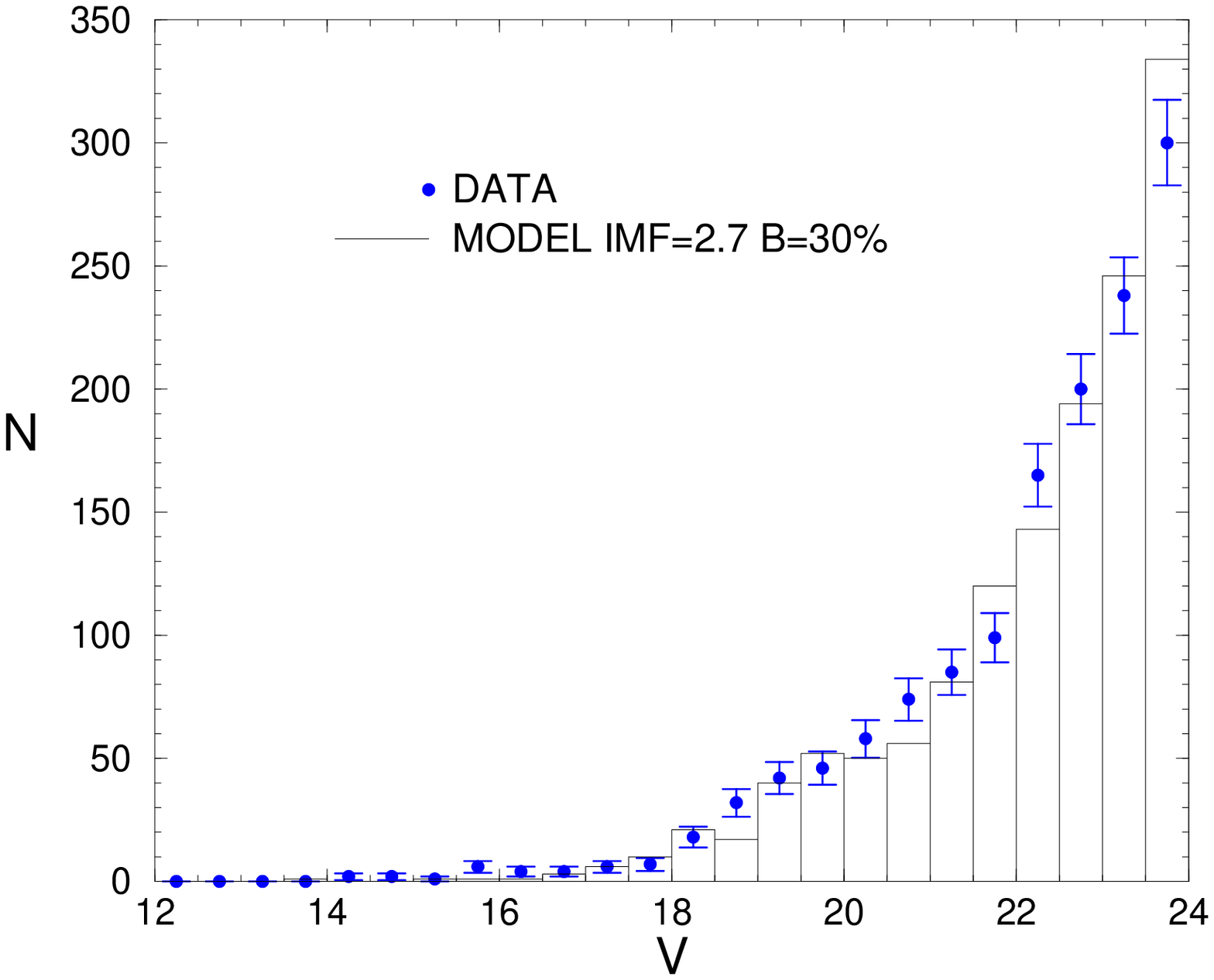}\\
\centering \includegraphics[width=6cm]{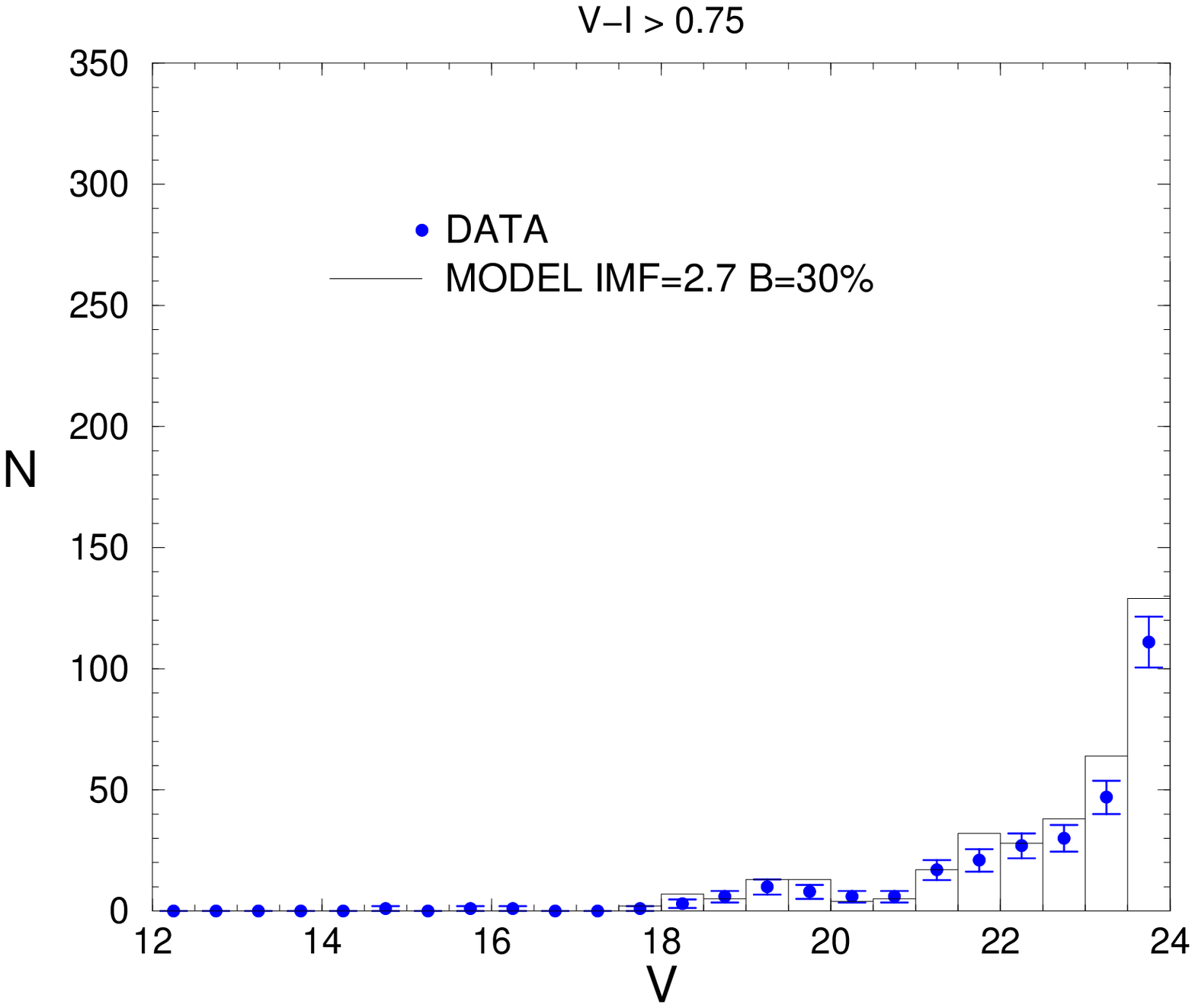}\\
\centering \includegraphics[width=6cm]{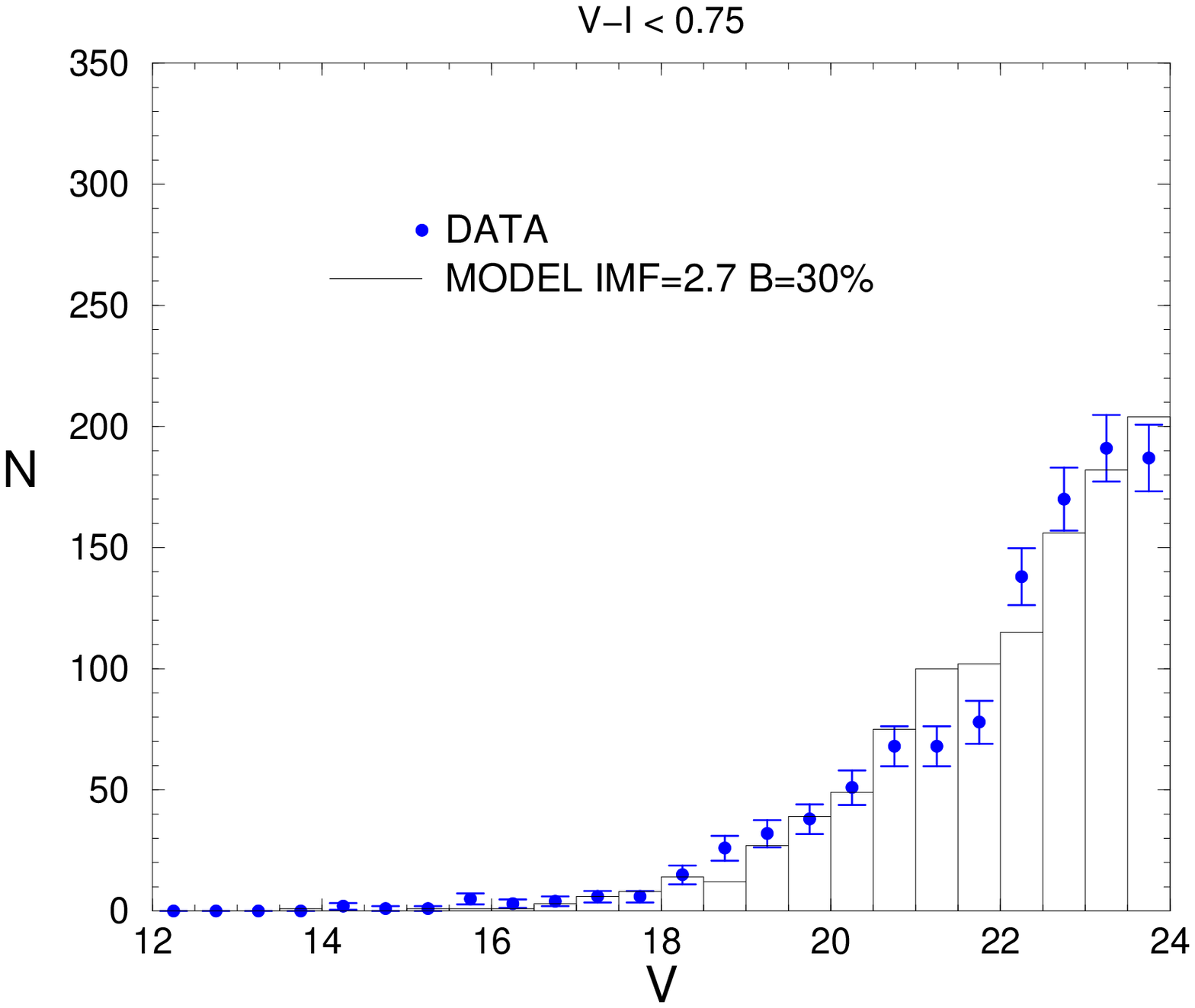}
\caption{The top panel shows the best synthetic diagram using an IMF exponent
$\alpha=2.7$. The remaining figures follow the same criteria of Figure
\ref{lfs1} but for $\alpha=2.7$.}
\label{lfs2} 
\end{figure}

\begin{figure}
\centering \includegraphics[width=6cm]{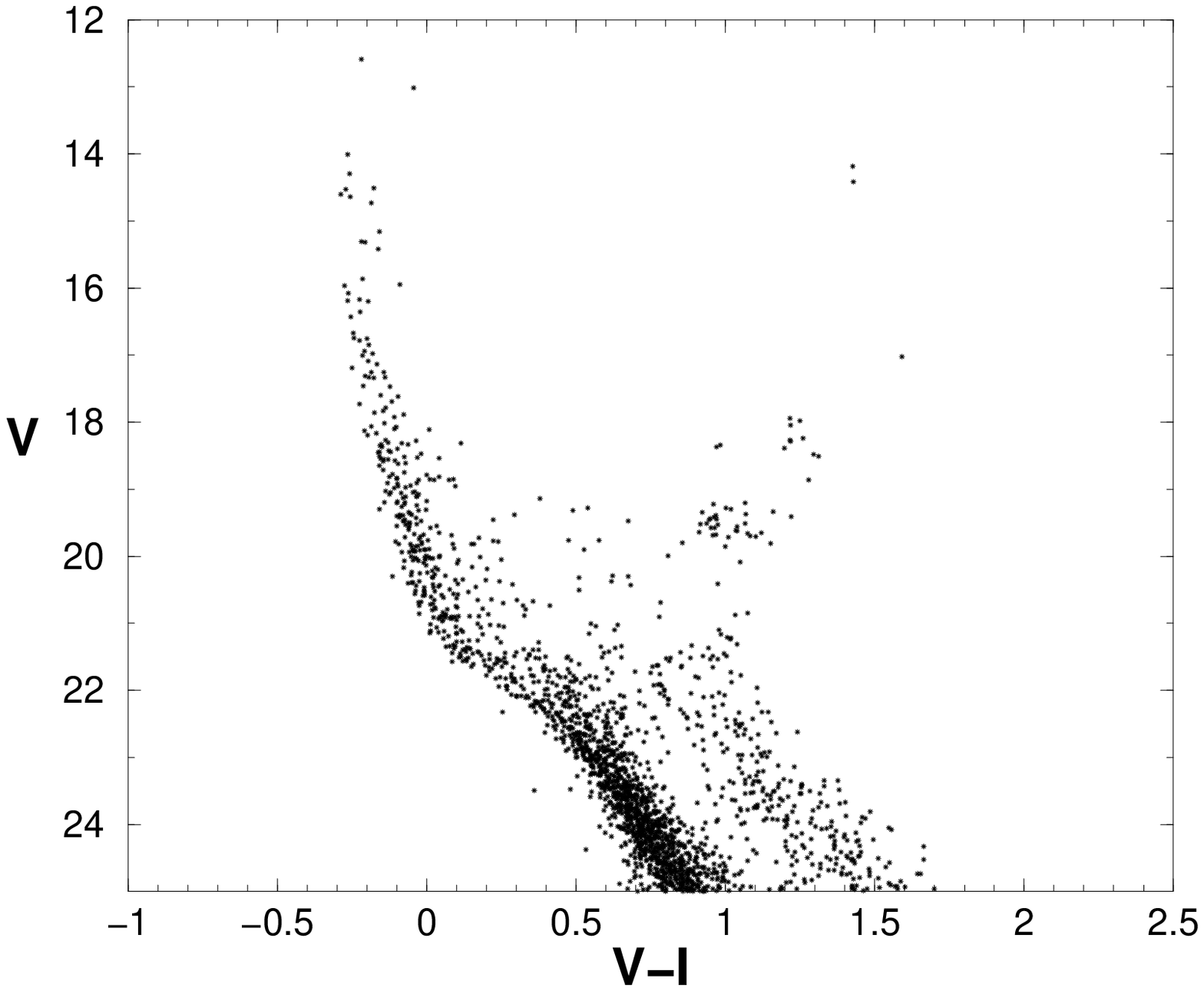}\\
\centering \includegraphics[width=6cm]{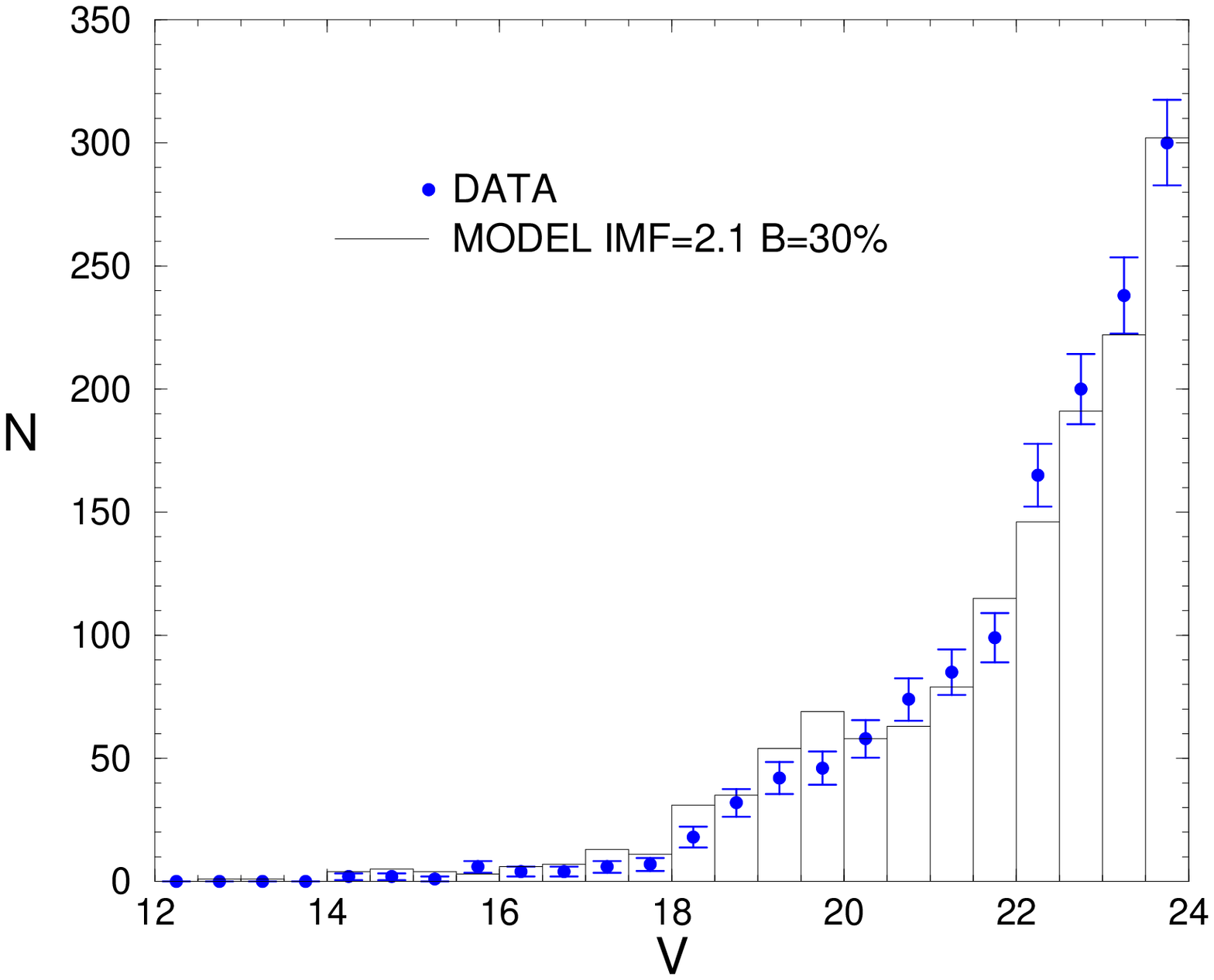}\\
\centering \includegraphics[width=6cm]{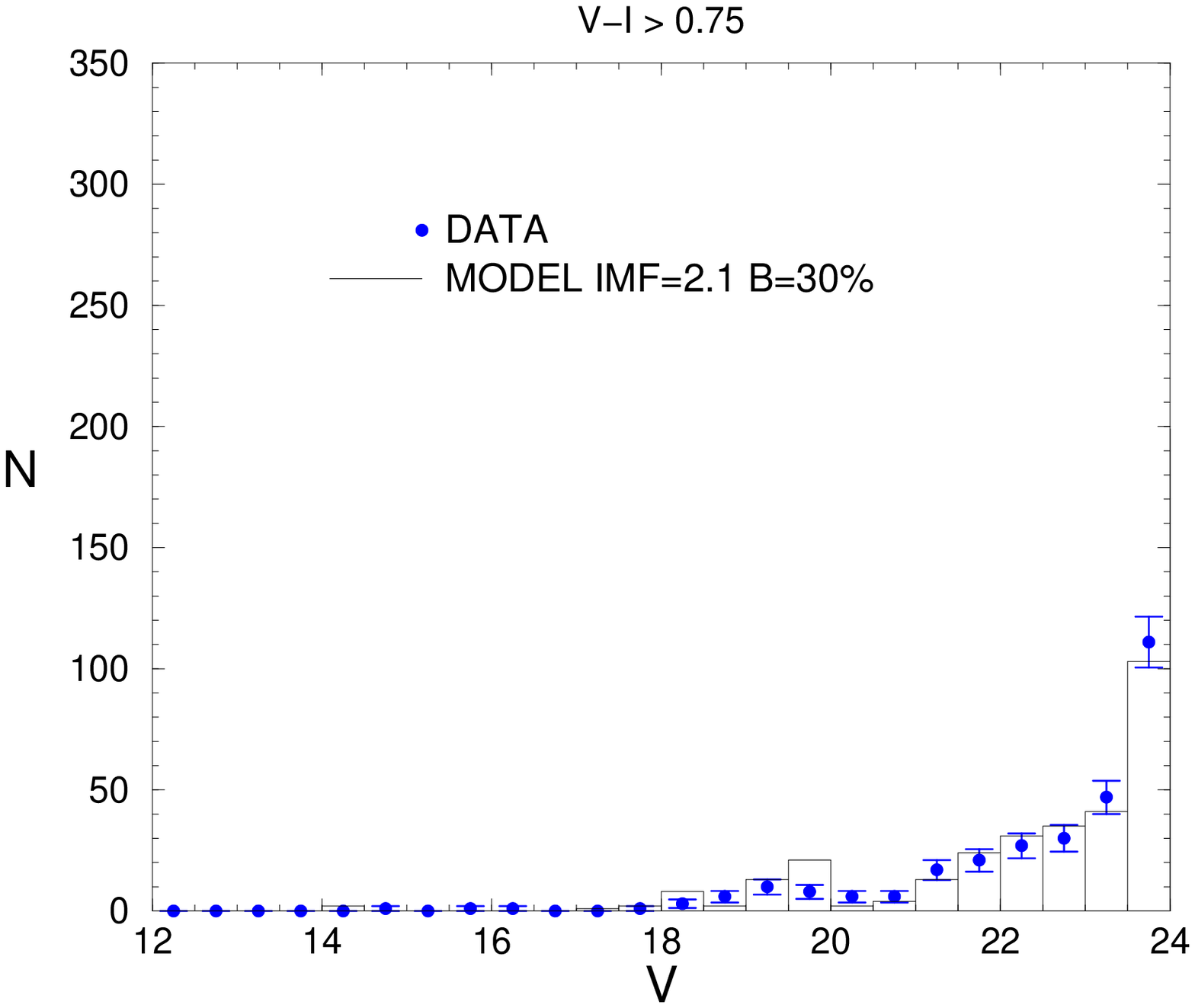}\\
\centering \includegraphics[width=6cm]{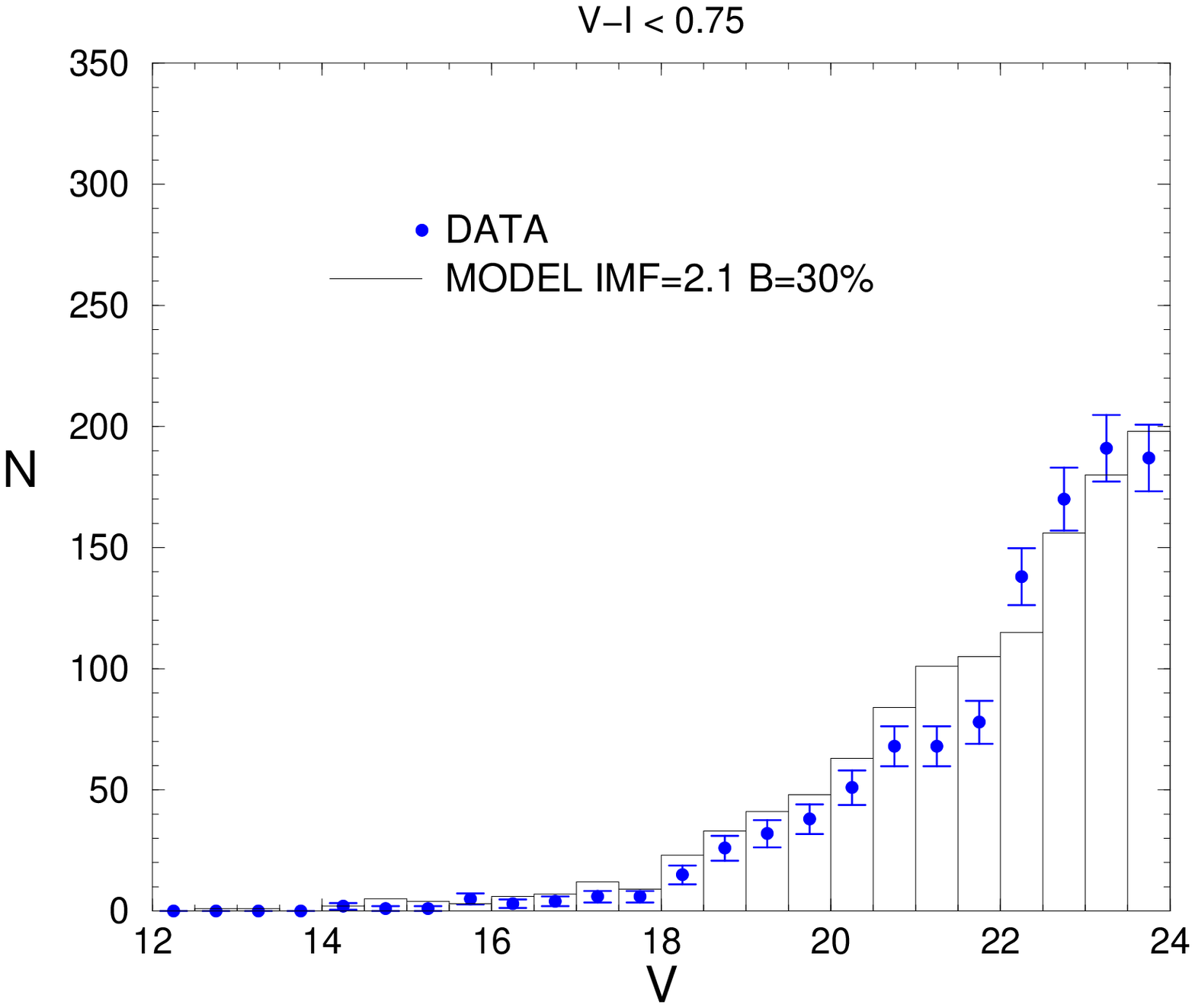}
\caption{As in Figure \ref{lfs2} but for $\alpha=2.1$.}
\label{lfs3} 
\end{figure}

Our results show that the luminosity functions built with an IMF exponent
$\alpha=2.5$ are nicely within the Poisson error bars (see
Fig. \ref{lfs1}). The small size of our sample makes it difficult to firmly
establish which value is best for the exponent.  Testing different $\alpha$'s,
however, we have found that only values between 2.1 and 2.7 are acceptable (see
Figs.  \ref{lfs2} and \ref{lfs3}).  For values higher than 2.7, the bright
main sequence population falls short of what is observed, while IMFs flatter
than $\alpha=2.1$ result in an over-populated upper main sequence.

\subsubsection{Binaries}
In this section we analyze the influence on the resulting SFH of the
assumed fraction of unresolved binaries. Partial CMDs have been
generated with 20\%, 30\%, 50\% or 70\% of unresolved binary systems
(for each binary, primary and secondary masses are both randomly
extracted from the same IMF). 

Concerning PMS stars, some observations seem to suggest a higher
binary fraction with respect to field stars. The Taurus star forming
region in the Milky Way has a factor of two excess of binaries,
although other regions show no excess. There is no consensus yet on
the origin of this phenomenon. Analyses of binary fractions in young
open clusters suggest that no evolutionary effect (leading to binary
disruption) is at work (at least after 50-70 Myr). Environmental
effects provide a more convincing explanation: the apparent
overabundance of young companions in Taurus with respect to the
solar neighborhood might imply that the SF in low-stellar-density
regions such as Taurus is particularly efficient to produce multiple
systems, and that such regions are not primary progenitors of our
solar neighborhood. However, the current sensitivity cannot exclude
that the discrepancy arise from an observational bias: the low mass
companions are brighter in PMS than in main sequence, making the
detection easier. Given these uncertainties, we prefer to adopt the
same binary fraction for all evolutionary phases.

\begin{figure}
\plotone{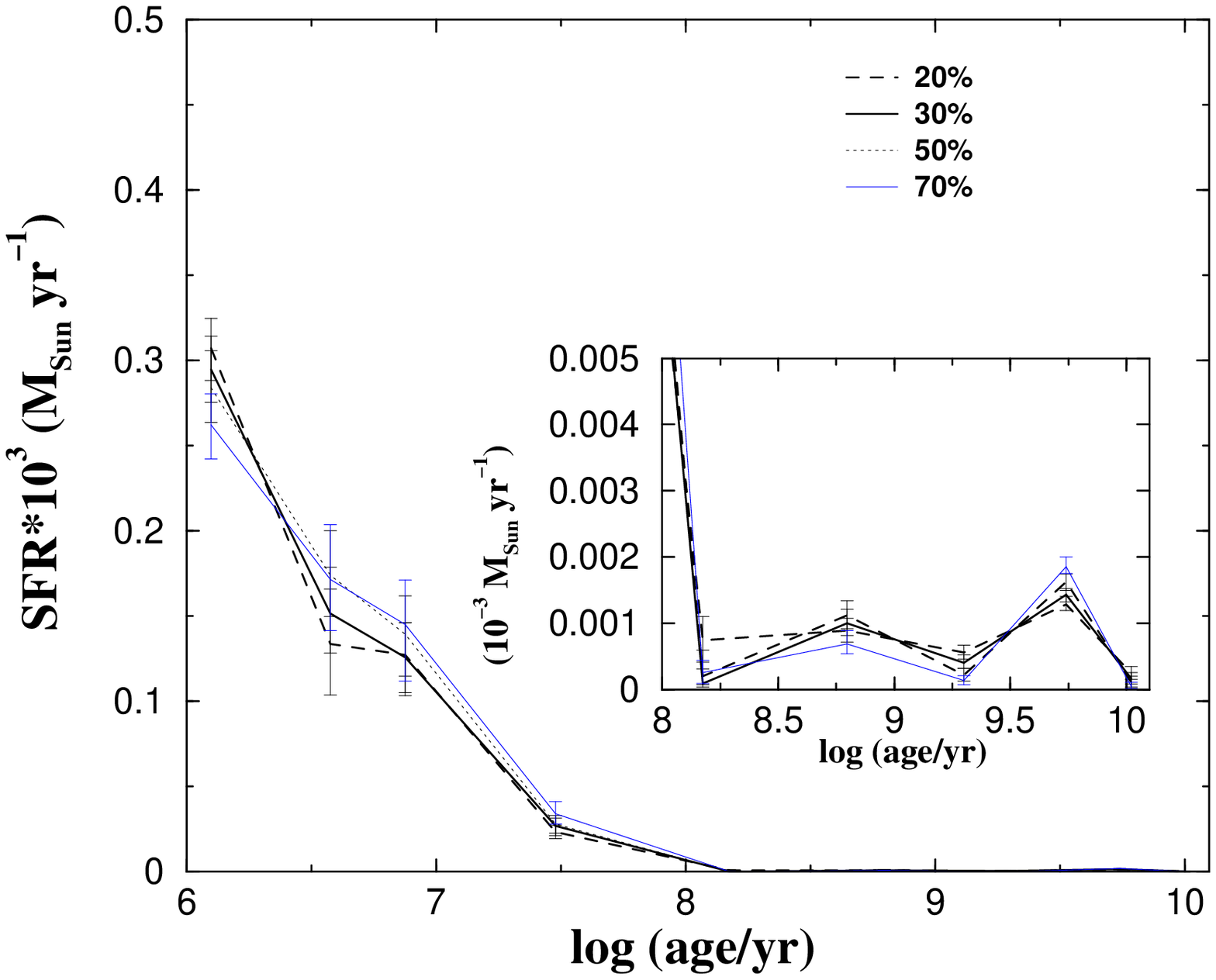}
\caption{Recovered star formation rate adopting different binary fractions (the
IMF exponent is fixed to 2.5).}
\label{figbin} 
\end{figure}

Figure \ref{figbin} shows the recovered SFHs: the overall effect of
increasing the binary fraction is a degradation of the information on
the SFH, washing out some features, both in recent and in past
epochs. This is because both the main sequence and the PMS
region are broadened by the presence of binaries, blending
contiguous temporal intervals.

Unlike the IMF, changing the fraction of binaries does not introduce a
systematic deviation; the effect is localized to specific
epochs. Binaries produce a secondary sequence redder and brighter than
the single star main sequence. When the stars evolve, they turn to the
red and in this sense a binary system may be confused with an evolved
single star. But, a binary system is also brighter, thus, it may be
also confused with a higher mass star (corresponding in main sequence
to a younger age). Even if the effect of binaries is quite complex,
our numerical results are comforting: the recovered SFH is not
significantly affected by this assumption.

In order to put some limits on the number of binaries in our field, the
synthetic CMDs are generated using the SFHs of Figure \ref{figbin} and the
luminosity functions are further explored.  Our analysis suggests that any
fraction lower than 70\% is equally acceptable because the photometric error
smears out the single and binary star sequences and makes them
indistinguishable from each other.

\begin{figure}
\centering \includegraphics[width=6cm]{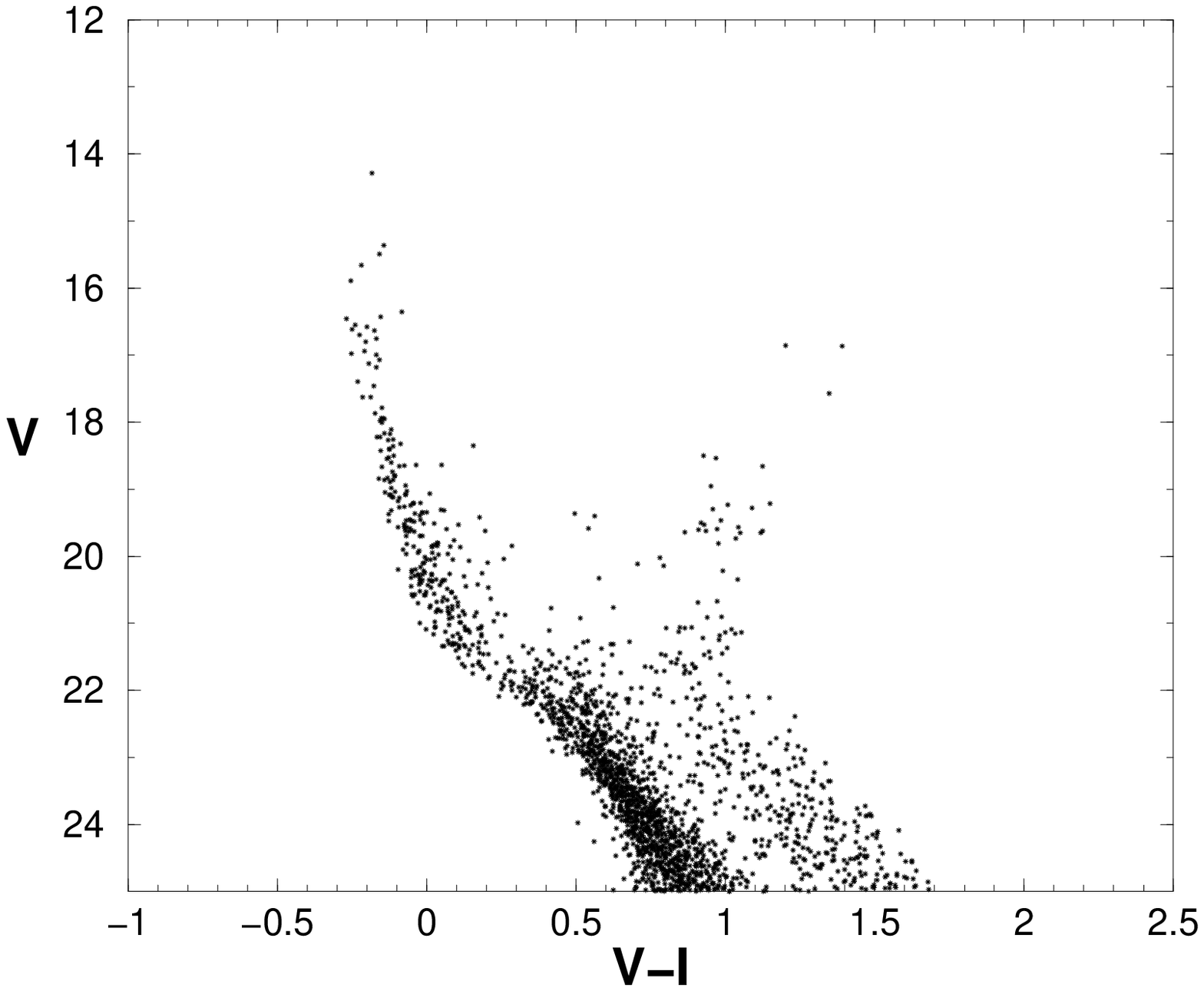}\\
\centering \includegraphics[width=6cm]{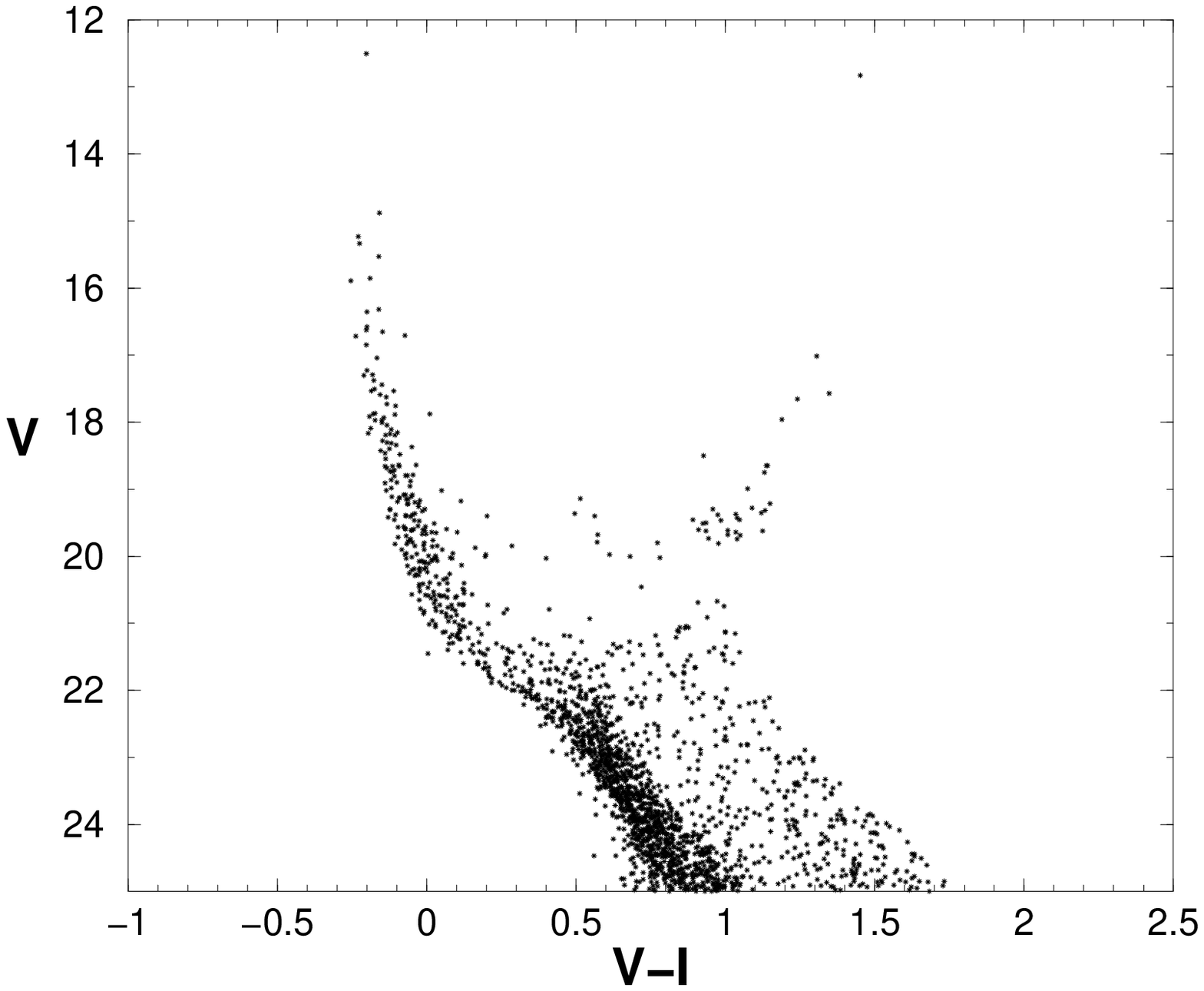}
\caption{Synthetic CMD: $\alpha=2.5$, binary fraction of 50\% (top
panel) and 70\% (bottom panel).}
\label{cmd50} 
\end{figure}

A visual inspection of the CMD allows additional considerations. Fig.
\ref{cmd50} (top panel) shows the CMD for a binary fraction of
50\%. Compared to the CMD generated with a 30\% (Fig. \ref{synth}),
the turn-off region around $V\sim 22$ appears broader, better matching
the observations. However, since the same color spread can be obtained
through many other ingredients (rotation, stronger overshooting,
reddening, metallicity, etc..), we do not see compelling evidence to
prefer this specific value.

To be conservative, only binary fractions larger than 70\% (bottom
panel) can be easily discarded. As visible in the bottom panel of
Fig. 22 the synthetic CMD assuming such large fractions lacks the
observed color separation between lower main sequence and PMS.

\subsubsection{Additional uncertainties regarding the PMS phase}

In addition to the ``classical'' uncertainties, some specific
issues are expected to affect only the PMS phase {\bf(see e.g. \citealt{sherry})}. First of all,
low-mass PMS stars are often identified as highly variable stars, so
called T Tauri-stars. Their large amplitude irregular variations in
the brightness are believed to arise from accretion shocks on the
stellar surface and/or chromospheric flaring. This effect, still not
well understood, can contribute to smearing the CMD in the PMS
region, mimicking an age spread. Differential reddening, which
preferentially affects stars still embedded in gas and dust like the
PMS stars, should also be considered as another plausible source of
confusion. Several studies have tried to evaluate the impact of
these uncertainties on the CMD. Studying OB associations like
$\sigma$ Orionis, \cite{burni2005} argue that the combination of
variability (on a typical period of a few years) and differential
reddening can produce, at most, an age spread of 4 Myr. This spread
is of the same order of magnitude as the result of \cite{henne} for
the NGC346 region (another SMC star forming region), who find that a
combination of binarity/variability/reddening might smear out any
SFH feature (as deduced from PMS alone) in the last 10 Myr. However,
these estimates may be an upper limit. In fact, there is a growing
evidence for large differences in the accretion disk (that is the
first candidate for variability and absorption in T-Tauri stars)
dissipation timescales for stars of a given mass in OB associations:
a fraction around 30\% appears to lack the disk at 1 Myr, while the
remainder hold accretion disks up to ages of $\sim 10-20$
Myr. Additional uncertainties are also the position of the
birth-line, which is strongly correlated with the number of T-Tauri,
and the treatment of convection, which is parameterized assuming a
mixing length parameter as derived from the Sun.

The lack of information about these uncertainties is so high that we do not attempt to include them in the models. Moreover, these uncertainties affect only the PMS, while we use all stellar phases to derive the SFH.

\section{Discussion and Conclusions}

In this paper, we presented a detailed look at the stellar content of an SMC
region (NGC~602), characterized by extremely low gas and dust density, but
displaying the properties of a very active and complex star forming
environment.

NGC~602 is located in the SMC ``wing'', and its stellar population is a
mixture of extremely young stars, born in the last few million years, and very
old stars, likely belonging to the SMC field population, of age 6-8 Gyr. The
young population is preferentially found in the densest central region.  A
rich population of PMS stars is observed, which appear still to be descending
their Hayashi tracks. This sets the maximum age for these stars at 5 Myr. In
addition, the central region of the cluster shows an excess of PMS and low
mass main sequence stars. If we add the compelling evidence \citep{carlson07}
that the star formation is currently ongoing along the outskirts of the
central cluster, where a significant number of YSOs have been observed in
Spitzer/IRAC images, the following scenario looks most attractive:
\begin{itemize}
\item an intense star formation activity began deep in the main cluster, NGC~602-A;
\item a less intense star formation process propagated or continued outside;
\item the activity has already stopped inside, while outside it is still
smoldering.
\end{itemize}

Using a metallicity $Z=0.004$ (consistent with the most recent determinations
for the SMC), we derive a distance modulus ${(m-M)_0}=18.9$ and a reddening
$E(B-V)=0.08$ for NGC~602. However, a visible red shift of the upper main
sequence with respect to the models could imply that the youngest stars are
affected by stronger reddening, and therefore the distance modulus could
decrease to ${(m-M)_0}=18.7$, consistent with the fact that NGC~602 is located
in the SMC ``wing''.

The old population is more uniformly distributed. Using a metallicity
$Z=0.001$ and the same distance/reddening combination as the young population,
we can simultaneously well reproduce the observed data for the lower sub-giant
branch and the helium clump color/luminosity. A higher metallicity would
require a much lower reddening, inconsistent with the Schlegel maps that
indicate $E(B-V)=0.06$.

With the objective of understanding how star formation started and progressed
in this region, we have used the detailed census of the stellar content to
derive the star formation history for NGC~602.

First of all, on the basis of the observational data, we estimated the
PDMF. Our results indicate that in the mass range between 0.7 and 30
M$_\odot$ the PDMF slope is $\Gamma$ = -1.25 $\pm$ 0.22 (see also
\citealt{schma}), and becomes $\Gamma$=-1.48 $\pm$ 0.17 if we
consider a 30\% fraction of unresolved binaries. We inferred a very
similar value for the IMF of NGC~602 ($\Gamma_{IMF}$=-1.5) using the
synthetic CMD method including binaries. Both our estimates are in
excellent agreement with the value derived by Salpeter in the solar
neighborhood. Therefore, despite the extremely low gas and dust
density, star cluster formation and evolution do not seem to be
significantly affected by the local conditions. We have used the
measured PDMF to derive the total mass in the NGC~602 region. By
integrating the observed PDMF between 0.7 and 120 M$_{\odot}$, we
calculate a total mass for NGC~602 of 1600 M$_{\odot}$, indicating
that the star formation event was quite modest in size.

\cite{nigra} have addressed the issue of what triggered the
initial episode of star formation in this region. They identified a distinct
HI cloud component that is likely the progenitor cloud of the cluster and
present HII region.  From HI survey data, they measured a HI density in
the natal cloud of 1.3 cm$^{-3}$ and a total mass of 3 $\times$ 10$^{5}$
M$_{\odot}$ based on a scale size of 220 pc.  This would imply an observed
star formation efficiency lower than 1\%, as typically expected.

We have used our population synthesis model to analyze the SFH of the
region. We have adopted the new PMS tracks for $Z=0.004$ computed with
the FRANEC evolutionary code, and presented in this paper. We have
assumed different choices for IMF and binary fraction, to test the
robustness of our results.  The best SFH is derived using a maximum
likelihood technique. We note that the derived SFHs turn out to be
qualitatively rather similar, independent of the assumptions on the
IMF or binary fraction, confirming the robustness of the result. The
star formation activity has been increasing with time on a time scale
of tens of Myr. The star formation in the recent 10 Myr has been quite
high, reaching a peak of $(0.3-0.7)\times$ 10$^{-3}$ M$_{\odot}$/yr
(the exact value depends on the IMF extrapolation) in the last 2.5
Myr. The current star formation is approximately 100 times higher than
the average from 4-8 Gyr ago. Any activity older than 8 Gyr ago seems to be negligible.

These results highlight the two different generations in the region: the
recent one, increasing in the near past and still ongoing, very spatially
concentrated, and the several Gyr old one, spatially diffused and related to the  SMC field.

We can assemble the results of \cite{nigra} and this paper to compose a
plausible scenario for the formation and evolution of this region. The SMC
experienced a first intense episode of SF in this region several Gyr ago
($\sim 6-8$), followed by a long period of moderate activity. Only
approximately 10 Myr ago, the compression and turbulence associated with the
interaction of two HI shells caused a progressive increase in the local star
formation rate, that reached $0.3 \times 10^{-3} M_\odot yr^{-1}$ in the
recent 2.5 Myr, and led to the formation of the central cluster
NGC~602-A. Star formation is still ongoing in the remaining reservoirs of
gas/dust at the outskirts of the cluster and it is likely to continue at the
same pace until the fuel is totally exhausted.

It is not straightforward to compare our rate with those of similar OB
associations. However, $(0.3-0.7)\times 10^{-3}\,\,M_{\odot}\,yr^{-1}$ seems in
good agreement with the results for several, relatively small scale, star
forming regions.  For example, \cite{prei} find for the Upper Scorpius
association in the Milky Way a star formation rate of
$10^{-3}\,\,M_{\odot}\,yr^{-1}$ on a projected radius of 10-15 pc. These stars
formed almost coevally about 5-6 Myr ago \citep{prei}.

A similar intensity in star formation is also found in the Orion Nebula:
\cite{hillenbrand98} report $10^{-4}-10^{-3}\,\,M_{\odot}\,yr^{-1}$ within a
projected radius of 2.5 pc. The mean age is younger than 6 Myr. However, while
the star forming activity in the Orion Nebula has stopped, the NGC~602 region
shows clear evidence of YSOs, a signature that the process has not completely
exhausted the fuel yet.  \cite{lada} find for NGC1333, an active region
(probably younger than 2 Myr) in the Perseus cloud, a star formation rate of
$4\times 10^{-5}\,\,M_{\odot}\,yr^{-1}$.

Although the NGC~602 star formation rate appears quite similar to other young
star clusters, it is much smaller than the stellar production in 30 Doradus,
a ``starburst-like'' complex in LMC: for this region, \cite{kenni} suggests
rates of $10^{-2}\,\,M_{\odot}\,yr^{-1}$ in a region of 10 pc, with a similar
age of ~3 Myr.


The remote location, the low density of gas and dust, and the quiescence of
the SMC wing are not favorable to star formation. Still NGC~602 is displaying
activity that is very comparable to known galactic young star clusters,
further confirming our understanding that local conditions have little effect
on the stellar content, for example, in terms of the distribution of masses
(MF). Remarkably our multi-wavelength approach allowed us to look back in
history at the very origin of this region, and follow the various phases of
its evolution.

The expectation is that for NGC~602 the exciting times are still to come when
the bulk of the massive stars will explode as SNeII and will significantly perturb the
local medium.

\section*{Acknowledgments}
We are grateful to P. Montegriffo for his software support, and to Howard Bond for acquiring the SMART data. L.A., M.C. and M.T. acknowledge financial support from PRIN-INAF-2005. Support for program \#10248 was provided by NASA to the US team members through a grant from the Space Telescope Science Institute, which is operated by AURA, Inc., under NASA contract NAS 5-26555. JSG also thanks the University of Wisconsin Graduate School for partial support of this research.



\begin{deluxetable}{ccc|cccc|cccc}
\tabletypesize{\tiny} \tablenum{1a} \tablewidth{0pt} \tablecaption{PMS tracks (Z=0.004, Y=0.24).}  \tablehead{ \colhead{Age (yr)} &
\colhead{$Log(L/L_{\odot})$} & \colhead{$Log(T_{eff})$} & & \colhead{Age(yr)}
& \colhead{$Log(L/L_{\odot})$} & \colhead{$Log(T_{eff})$} & &
\colhead{Age(yr)} & \colhead{$Log(L/L_{\odot})$}&\colhead{$Log(T_{eff})$}}
\startdata &&&&&&&&&&\\
&$0.45\,M_{\odot}$&&&&$0.50\,M_{\odot}$&&&&$0.60\,M_{\odot}$&\\ &&&&&&&&&&\\
\tableline 
0.10E+05 & 0.337  & 3.618&&0.10E+05 & 0.475  & 3.629&&0.10E+05 & 0.694  & 3.645\\ 
0.20E+05 & 0.316  & 3.617&&0.20E+05 & 0.449  & 3.629&&0.20E+05 & 0.661  & 3.645\\ 
0.30E+05 & 0.295  & 3.617&&0.30E+05 & 0.425  & 3.629&&0.30E+05 & 0.630  & 3.644\\ 
0.50E+05 & 0.258  & 3.616&&0.50E+05 & 0.381  & 3.628&&0.50E+05 & 0.577  & 3.644\\ 
0.70E+05 & 0.224  & 3.615&&0.70E+05 & 0.343  & 3.627&&0.70E+05 & 0.531  & 3.644\\ 
0.10E+06 & 0.179  & 3.614&&0.10E+06 & 0.292  & 3.625&&0.10E+06 & 0.473  & 3.643\\ 
0.20E+06 & 0.065  & 3.610&&0.20E+06 & 0.164  & 3.621&&0.20E+06 & 0.332  & 3.640\\ 
0.30E+06 & -0.018 & 3.607&&0.30E+06 & 0.074  & 3.618&&0.30E+06 & 0.234  & 3.637\\ 
0.40E+06 & -0.082 & 3.605&&0.40E+06 & 0.005  & 3.615&&0.40E+06 & 0.158  & 3.634\\ 
0.50E+06 & -0.135 & 3.604&&0.50E+06 & -0.051 & 3.613&&0.50E+06 & 0.094  & 3.631\\ 
0.60E+06 & -0.179 & 3.603&&0.60E+06 & -0.098 & 3.612&&0.60E+06 & 0.043  & 3.629\\ 
0.70E+06 & -0.218 & 3.602&&0.70E+06 & -0.139 & 3.610&&0.70E+06 & -0.001 & 3.628\\ 
0.80E+06 & -0.252 & 3.601&&0.80E+06 & -0.175 & 3.609&&0.80E+06 & -0.039 & 3.626\\ 
0.10E+07 & -0.310 & 3.600&&0.10E+07 & -0.235 & 3.608&&0.10E+07 & -0.104 & 3.623\\
0.20E+07 & -0.495 & 3.597&&0.20E+07 & -0.425 & 3.603&&0.20E+07 & -0.304 & 3.617\\ 
0.30E+07 & -0.605 & 3.596&&0.30E+07 & -0.538 & 3.601&&0.30E+07 & -0.420 & 3.614\\ 
0.50E+07 & -0.748 & 3.594&&0.50E+07 & -0.682 & 3.600&&0.50E+07 & -0.565 & 3.611\\ 
0.70E+07 & -0.844 & 3.593&&0.70E+07 & -0.777 & 3.599&&0.70E+07 & -0.658 & 3.610\\ 
0.10E+08 & -0.945 & 3.592&&0.10E+08 & -0.876 & 3.598&&0.10E+08 & -0.750 & 3.609\\ 
0.20E+08 & -1.136 & 3.592&&0.20E+08 & -1.057 & 3.598&&0.20E+08 & -0.891 & 3.614\\
0.30E+08 & -1.237 & 3.592&&0.30E+08 & -1.142 & 3.601&&0.30E+08 & -0.924 & 3.623\\ 
0.50E+08 & -1.336 & 3.595&&0.50E+08 & -1.209 & 3.608&&0.50E+08 & -0.845 & 3.656\\ 
0.10E+09 & -1.394 & 3.604&&0.10E+09 & -1.238 & 3.620&&0.10E+09 & -0.970 & 3.647\\ 
0.18E+09 & -1.437 & 3.601&&0.15E+09 & -1.288 & 3.613&&0.11E+09 & -0.968 & 3.647\\ 
\tableline 
&&&&&&&&&&\\
&$0.70\,M_{\odot}$&&&&$0.80\,M_{\odot}$&&&&$0.90\,M_{\odot}$&\\ 
&&&&&&&&&&\\
\tableline 
0.10E+05 & 0.868  & 3.655&& 0.10E+05 &  1.007 & 3.662&& 0.10E+05 &  1.116 & 3.667\\ 
0.20E+05 & 0.827  & 3.655&& 0.20E+05 &  0.960 & 3.662&& 0.20E+05 &  1.065 & 3.668\\ 
0.30E+05 & 0.791  & 3.655&& 0.30E+05 &  0.919 & 3.663&& 0.30E+05 &  1.021 & 3.669\\ 
0.50E+05 & 0.730  & 3.655&& 0.50E+05 &  0.852 & 3.664&& 0.50E+05 &  0.949 & 3.671\\ 
0.70E+05 & 0.678  & 3.655&& 0.70E+05 &  0.796 & 3.664&& 0.70E+05 &  0.891 & 3.671\\ 
0.10E+06 & 0.615  & 3.655&& 0.10E+06 &  0.728 & 3.665&& 0.10E+06 &  0.821 & 3.672\\ 
0.20E+06 & 0.464  & 3.654&& 0.20E+06 &  0.572 & 3.664&& 0.20E+06 &  0.661 & 3.673\\ 
0.30E+06 & 0.362  & 3.652&& 0.30E+06 &  0.468 & 3.663&& 0.30E+06 &  0.557 & 3.672\\ 
0.40E+06 & 0.285  & 3.650&& 0.40E+06 &  0.389 & 3.662&& 0.40E+06 &  0.478 & 3.671\\ 
0.50E+06 & 0.222  & 3.648&& 0.50E+06 &  0.326 & 3.660&& 0.50E+06 &  0.415 & 3.670\\ 
0.60E+06 & 0.169  & 3.646&& 0.60E+06 &  0.274 & 3.659&& 0.60E+06 &  0.363 & 3.670\\ 
0.70E+06 & 0.124  & 3.645&& 0.70E+06 &  0.228 & 3.658&& 0.70E+06 &  0.317 & 3.669\\
0.80E+06 & 0.084  & 3.643&& 0.80E+06 &  0.188 & 3.656&& 0.80E+06 &  0.277 & 3.668\\ 
0.10E+07 & 0.017  & 3.640&& 0.10E+07 &  0.120 & 3.654&& 0.10E+07 &  0.210 & 3.666\\ 
0.20E+07 & -0.199 & 3.630&& 0.20E+07 & -0.096 & 3.645&& 0.20E+07 & -0.002 & 3.658\\ 
0.30E+07 & -0.320 & 3.625&& 0.30E+07 & -0.222 & 3.639&& 0.30E+07 & -0.119 & 3.654\\ 
0.50E+07 & -0.462 & 3.622&& 0.50E+07 & -0.358 & 3.635&& 0.50E+07 & -0.239 & 3.654\\ 
0.70E+07 & -0.547 & 3.621&& 0.70E+07 & -0.428 & 3.636&& 0.70E+07 & -0.283 & 3.660\\ 
0.10E+08 & -0.624 & 3.622&& 0.10E+08 & -0.475 & 3.642&& 0.10E+08 & -0.274 & 3.677\\ 
0.20E+08 & -0.681 & 3.639&& 0.20E+08 & -0.359 & 3.698&& 0.20E+08 & -0.021 & 3.759\\ 
0.30E+08 & -0.565 & 3.682&& 0.30E+08 & -0.224 & 3.752&& 0.30E+08 & -0.142 & 3.776\\
0.50E+08 & -0.607 & 3.709&& 0.44E+08 & -0.395 & 3.742&& 0.35E+08 & -0.163 & 3.773\\ 
0.74E+08 & -0.654 & 3.700&&          &        &      &&          &        &      \\ 
\tableline 
&&&&&&&&&&\\
&$1.00\,M_{\odot}$&&&&$1.20\,M_{\odot}$&&&&$1.50\,M_{\odot}$&\\ 
&&&&&&&&&&\\
\tableline 
0.10E+05 & 1.224  & 3.671&& 0.10E+05 & 1.330   & 3.679&&0.10E+05 & 1.585    & 3.683\\ 
0.20E+05 & 1.167  & 3.673&& 0.20E+05 & 1.275   & 3.681&&0.20E+05 & 1.507    & 3.686\\ 
0.30E+05 & 1.119  & 3.674&& 0.30E+05 & 1.229   & 3.682&&0.30E+05 & 1.447    & 3.688\\ 
0.50E+05 & 1.041  & 3.676&& 0.50E+05 & 1.154   & 3.685&&0.50E+05 & 1.354    & 3.692\\ 
0.70E+05 & 0.980  & 3.677&& 0.70E+05 & 1.095   & 3.686&&0.70E+05 & 1.285    & 3.694\\ 
0.10E+06 & 0.906  & 3.678&& 0.10E+06 & 1.024   & 3.688&&0.10E+06 & 1.204    & 3.697\\ 
0.20E+06 & 0.742  & 3.680&& 0.20E+06 & 0.864   & 3.690&&0.20E+06 & 1.032    & 3.701\\ 
0.30E+06 & 0.636  & 3.680&& 0.30E+06 & 0.760   & 3.691&&0.30E+06 & 0.924    & 3.703\\ 
0.40E+06 & 0.557  & 3.679&& 0.40E+06 & 0.683   & 3.692&&0.40E+06 & 0.846    & 3.705\\ 
0.50E+06 & 0.494  & 3.679&& 0.50E+06 & 0.621   & 3.692&&0.50E+06 & 0.784    & 3.705\\ 
0.60E+06 & 0.442  & 3.678&& 0.60E+06 & 0.569   & 3.692&&0.60E+06 & 0.735    & 3.706\\
0.70E+06 & 0.396  & 3.678&& 0.70E+06 & 0.526   & 3.692&&0.70E+06 & 0.693    & 3.707\\ 
0.80E+06 & 0.357  & 3.677&& 0.80E+06 & 0.488   & 3.691&&0.80E+06 & 0.658    & 3.707\\ 
0.10E+07 & 0.290  & 3.675&& 0.10E+07 & 0.424   & 3.691&&0.10E+07 & 0.602    & 3.708\\ 
0.20E+07 & 0.085  & 3.670&& 0.20E+07 & 0.244   & 3.691&&0.20E+07 & 0.476    & 3.718\\ 
0.30E+07 & -0.022 & 3.668&& 0.30E+07 & 0.170   & 3.695&&0.30E+07 & 0.485    & 3.734\\ 
0.50E+07 & -0.112 & 3.674&& 0.50E+07 & 0.166   & 3.716&&0.50E+07 & 0.782    & 3.782\\ 
0.70E+07 & -0.119 & 3.687&& 0.70E+07 & 0.275   & 3.745&&0.70E+07 & 1.022    & 3.888\\ 
0.10E+08 & -0.045 & 3.717&& 0.10E+08 & 0.530   & 3.794&&0.10E+08 & 0.861    & 3.934\\ 
0.20E+08 &  0.191 & 3.804&& 0.16E+08 & 0.417   & 3.843&&0.12E+08 & 0.850    & 3.934\\ 
0.27E+08 &  0.046 & 3.797&&          &         &      &&         &          &      \\
%
%
\tableline
&&&&&&&&&&\\
&$2.00\,M_{\odot}$&&&&$3.00\,M_{\odot}$&&&&$4.00\,M_{\odot}$&\\
&&&&&&&&&&\\
\tableline
0.10E+05 &  1.815 &  3.689&& 0.10E+05 &  2.122 &  3.695&& 0.10E+05 &  2.329 &  3.697\\ 
0.20E+05 &  1.726 &  3.693&& 0.20E+05 &  2.016 &  3.701&& 0.20E+05 &  2.219 &  3.705\\  
0.30E+05 &  1.658 &  3.696&& 0.30E+05 &  1.941 &  3.705&& 0.30E+05 &  2.141 &  3.710\\  
0.50E+05 &  1.558 &  3.701&& 0.50E+05 &  1.832 &  3.710&& 0.50E+05 &  2.037 &  3.716\\  
0.70E+05 &  1.485 &  3.704&& 0.70E+05 &  1.756 &  3.714&& 0.70E+05 &  1.970 &  3.721\\  
0.10E+06 &  1.401 &  3.707&& 0.10E+06 &  1.672 &  3.718&& 0.10E+06 &  1.908 &  3.726\\  
0.20E+06 &  1.226 &  3.713&& 0.20E+06 &  1.516 &  3.727&& 0.20E+06 &  1.892 &  3.742\\  
0.30E+06 &  1.121 &  3.716&& 0.30E+06 &  1.444 &  3.734&& 0.30E+06 &  2.054 &  3.768\\  
0.40E+06 &  1.048 &  3.719&& 0.40E+06 &  1.416 &  3.739&& 0.40E+06 &  2.222 &  3.845\\  
0.50E+06 &  0.994 &  3.721&& 0.50E+06 &  1.439 &  3.746&& 0.50E+06 &  2.414 &  3.956\\  
0.60E+06 &  0.953 &  3.723&& 0.60E+06 &  1.519 &  3.756&& 0.60E+06 &  2.562 &  4.068\\  
0.70E+06 &  0.922 &  3.725&& 0.70E+06 &  1.630 &  3.768&& 0.70E+06 &  2.644 &  4.167\\  
0.80E+06 &  0.898 &  3.727&& 0.80E+06 &  1.720 &  3.785&& 0.80E+06 &  2.626 &  4.230\\  
0.10E+07 &  0.867 &  3.731&& 0.10E+07 &  1.910 &  3.869&& 0.10E+07 &  2.456 &  4.237\\  
0.20E+07 &  1.023 &  3.764&& 0.20E+07 &  2.016 &  4.163&& 0.11E+07 &  2.445 &  4.234\\  
0.30E+07 &  1.380 &  3.856&& 0.21E+07 &  2.002 &  4.160&&          &        &       \\ 
0.50E+07 &  1.388 &  4.045&&          &        &       &&          &        &       \\
0.61E+07 &  1.349 &  4.041&&          &        &       &&          &        &       \\
\tableline
&&&&&&&&&&\\
&$5.00\,M_{\odot}$&&&&$5.50\,M_{\odot}$&&&&&\\
&&&&&&&&&&\\
\tableline
0.10E+05 &  2.498 &  3.700&& 0.10E+05 &  2.556 &  3.700&&  &&\\
0.20E+05 &  2.382 &  3.707&& 0.20E+05 &  2.460 &  3.709&&  &&\\
0.30E+05 &  2.313 &  3.712&& 0.30E+05 &  2.406 &  3.714&&  &&\\
0.50E+05 &  2.248 &  3.722&& 0.50E+05 &  2.395 &  3.725&&  &&\\
0.70E+05 &  2.233 &  3.727&& 0.70E+05 &  2.487 &  3.736&&  &&\\
0.10E+06 &  2.315 &  3.740&& 0.10E+06 &  2.551 &  3.766&&  &&\\
0.20E+06 &  2.536 &  3.862&& 0.20E+06 &  2.893 &  4.006&&  &&\\
0.30E+06 &  2.843 &  4.060&& 0.30E+06 &  3.111 &  4.231&&  &&\\
0.40E+06 &  2.979 &  4.229&& 0.40E+06 &  3.042 &  4.325&&  &&\\
0.50E+06 &  2.913 &  4.302&& 0.49E+06 &  2.920 &  4.312&&  &&\\
0.62E+06 &  2.779 &  4.289&&          &        &       &&  &&\\
\enddata
\label{tab1} 
\end{deluxetable}

\end{document}